\documentclass[prd,preprint,nofootinbib]{revtex4-1}

\usepackage{amssymb,amsmath}
\usepackage{color}
\usepackage{graphicx,subfigure}
\usepackage{hyperref}
\usepackage{float} 
\usepackage[T1]{fontenc}

\newcommand{\nn}{\nonumber}
\newcommand{\A}{\mathcal{A}}

\newcommand \ket[1]{
        \left| #1 \right>
}

\newcommand \bra[1]{
        \left< #1 \right|
}

\begin{document}

\title{The Emergence of the $\Delta U=0$ Rule in Charm Physics}

\author{Yuval Grossman}
\email{yg73@cornell.edu}
\affiliation{Department of Physics, LEPP, Cornell University, Ithaca, NY 14853, USA}
\author{Stefan Schacht}
\email{ss3843@cornell.edu}
\affiliation{Department of Physics, LEPP, Cornell University, Ithaca, NY 14853, USA}

\begin{abstract}
We discuss the implications of the 
recent discovery of CP violation in two-body SCS $D$ decays by LHCb.
We show that the result can be explained within the SM without the need
for any large $SU(3)$ breaking effects. It further
enables the determination of the imaginary part of the ratio of the $\Delta U=0$ over $\Delta U=1$ matrix elements in charm decays,
which we find to be $(0.65\pm 0.12)$.
Within the standard model, the result proves the non-perturbative nature of the penguin contraction of 
tree operators in charm decays, similar to the known non-perturbative enhancement of $\Delta I=1/2$ over $\Delta I=3/2$ 
matrix elements in kaon decays, that is, the $\Delta I=1/2$ rule.
As a guideline for future measurements, we show how to completely
solve the most general parametrization of the $D \to P^+P^-$ system.
\end{abstract}

\maketitle

\section{Introduction \label{sec:intro}}

In a recent spectacular result, LHCb discovered direct CP violation in
charm decays at~5.3$\sigma$~\cite{Aaij:2019kcg}. 
The new world average of the difference of 
CP asymmetries \cite{Aitala:1997ff, Link:2000aw, Csorna:2001ww, Aubert:2007if, Staric:2008rx, Aaltonen:2011se, Collaboration:2012qw, Aaij:2011in, Aaij:2013bra, Aaij:2014gsa, Aaij:2016cfh, Aaij:2016dfb}
\begin{align}
\Delta a_{CP}^{\mathrm{dir}} &\equiv 
a_{CP}^{\mathrm{dir}}(D^0\rightarrow K^+K^-) - a_{CP}^{\mathrm{dir}}(D^0\rightarrow \pi^+\pi^-)\,, 
\end{align}
where 
\begin{align}
a_{CP}^{\mathrm{dir}}(f) &\equiv \frac{
\vert \mathcal{A} (D^0\to f)\vert^2 - \vert {\mathcal{A}}(\overline{D}^0\to f)\vert^2 
}{
\vert \mathcal{A}(D^0\to f)\vert^2 + \vert {\mathcal{A}}(\overline{D}^0\to f)\vert^2 
}\,, 
\end{align}
and which is provided by the Heavy Flavor Averaging Group (HFLAV) \cite{Amhis:2016xyh}, is given as~\cite{Carbone:2019}
\begin{align}
\Delta a_{CP}^{\mathrm{dir}} &= -0.00164\pm  0.00028\,. \label{eq:HFLAVav} 
\end{align}
Our aim in this paper is to study the implications of this result. In
particular, working within the Standard Model (SM) and using the known values of the Cabibbo-Kobayashi-Maskawa (CKM) matrix elements as input, we see how 
Eq.~(\ref{eq:HFLAVav}) can be employed in order to extract low energy QCD quantities,
and learn from them about QCD.

The new measurement allows for the first time to determine the CKM-suppressed amplitude of singly-Cabibbo-suppressed (SCS)
charm decays that contribute a weak phase difference relative to the CKM-leading part, which leads to a non-vanishing 
CP asymmetry. 
More specifically, $\Delta a_{CP}^{\mathrm{dir}}$ allows to determine
the imaginary part of the $\Delta U=0$ over $\Delta U=1$ matrix elements.

As we show, the data suggest the emergence of a $\Delta U=0$ rule, which has
features that are similar to the known
\lq\lq{}$\Delta I=1/2$ rule\rq\rq{} 
 in kaon physics.
This rule is the 
observation that in $K \to \pi\pi$ the amplitude into a $I=0$ final state is 
enhanced by a factor $\sim 20$ with respect to the one into a $I=2$ final 
state~\cite{Tanabashi:2018oca, GellMann:1955jx, GellMann:1957wh, Gaillard:1974nj, Bardeen:1986vz, Buras:2014maa, Bai:2015nea, Blum:2015ywa,Boyle:2012ys, Buras:2015yba, Kitahara:2016nld}. This is explained by
large non-perturbative rescattering effects. 
Analogous enhancements in charm decays have previously been discussed in 
Refs.~\cite{Einhorn:1975fw, Abbott:1979fw,Golden:1989qx,  Brod:2012ud,
  Grinstein:2014aza, Bhattacharya:2012ah, Franco:2012ck, Hiller:2012xm}.
For further recent theoretical work on charm CP violation see 
Refs.~\cite{Nierste:2017cua, Nierste:2015zra, Muller:2015rna, Grossman:2018ptn, Buccella:1994nf, Grossman:2006jg, Artuso:2008vf, Khodjamirian:2017zdu, Buccella:2013tya, Cheng:2012wr, Feldmann:2012js, Li:2012cfa, Atwood:2012ac, Grossman:2012ry, Buccella:2019kpn, Yu:2017oky, Brod:2011re}. 

In Sec.~\ref{sec:decomposition} we review the completely general U-spin decomposition of the decays 
$D^0\rightarrow K^+K^-$, $D^0\rightarrow \pi^+\pi^-$ and $D^0\rightarrow K^{\pm}\pi^{\mp}$. 
After that, in Sec.~\ref{sec:solving} we show how to completely determine all U-spin parameters from data. 
Our numerical results which are based on the current measurements are given in Sec.~\ref{sec:numerics}.
In Sec.~\ref{sec:deltau0rule} we interpret these as the emergence of a
$\Delta U=0$ rule, and in Sec.~\ref{sec:DeltaI12inKDB} we compare it to
the $\Delta I=1/2$ rules in $K$, $B$ and $D$ decays. The different effect of $\Delta U=0$ and $\Delta I=1/2$ rules
on the phenomenology of charm and kaon decays, respectively, is discussed in Sec.~\ref{sec:UandIrules}. 
In Sec.~\ref{sec:conclusions} we conclude.

\section{Most general amplitude decomposition \label{sec:decomposition}}

The Hamiltonian of SCS decays can be written as the sum 
\begin{align}
\mathcal{H}_{\mathrm{eff}} \sim \Sigma (1,0) - \frac{\lambda_b}{2} (0,0)\,,
\end{align}
where $(i,j) = \mathcal{O}^{\Delta U=i}_{\Delta U_3=j}$,
and the appearing combination of CKM matrix elements are
\begin{align}
\Sigma        &\equiv \frac{V_{cs}^* V_{us} - V_{cd}^* V_{ud}}{2}\,, \qquad 
-\frac{\lambda_b}{2}  \equiv -\frac{V_{cb}^* V_{ub}}{2} = \frac{V_{cs}^* V_{us} + V_{cd}^* V_{ud} }{2}\,, 
\end{align}
where numerically, $|\Sigma| \gg |\lambda_b|$. 
The corresponding amplitudes have the structure
\begin{align}
\mathcal{A} = \Sigma ( A_{\Sigma}^s - A_{\Sigma}^d ) - \frac{\lambda_b}{2} A_b\,,
\end{align}
where $A_{\Sigma}^s$, $A_{\Sigma}^d$ and $A_b$ contain only strong phases and we 
write also $A_{\Sigma}\equiv A_{\Sigma}^s - A_{\Sigma}^d$\,.

For the amplitudes we use the notation
\begin{align}
\mathcal{A}(K\pi)   &\equiv \mathcal{A}(\overline{D}^0 \rightarrow K^+\pi^-)\,, \\
\mathcal{A}(\pi\pi) &\equiv \mathcal{A}(\overline{D}^0 \rightarrow \pi^+\pi^-)\,, \\
\mathcal{A}(KK)     &\equiv \mathcal{A}(\overline{D}^0 \rightarrow K^+K^-)\,, \\
\mathcal{A}(\pi K)  &\equiv \mathcal{A}(\overline{D}^0 \rightarrow \pi^+ K^-)\,.
\end{align}
The U-spin related quartet of charm meson decays into charged final states can then be 
written as~\cite{Brod:2012ud,Muller:2015lua,Muller:2015rna}
\begin{align}
\A(K\pi)   &= V_{cs} V_{ud}^* \left(t_0 - \frac{1}{2} t_1  \right)\,, \label{eq:decomp-1}\\
\A(\pi\pi) &= -\Sigma^*\left(t_0 + s_1  + \frac{1}{2} t_2  \right)  
-\lambda_b^*\left(p_0 - \frac{1}{2} p_1  \right)\,, \label{eq:decomp-2}\\
\A(KK)     &= \Sigma^*\left(t_0 - s_1  + \frac{1}{2} t_2  \right)  
       -\lambda_b^* \left(p_0 + \frac{1}{2} p_1  \right)\,,  \label{eq:decomp-3}\\
\A(\pi K)   &= V_{cd} V_{us}^* \left(t_0 + \frac{1}{2} t_1  \right)\,. \label{eq:decomp-4}
\end{align}
The subscript of the parameters denotes the level of U-spin breaking
at which they enter.  
We write $A(K\pi)$ and $A(\pi K)$ for the Cabibbo-favored (CF) and doubly Cabibbo-suppressed (DCS) amplitude 
without the CKM factors, respectively.
We emphasize that the SM parametrization in Eqs.~(\ref{eq:decomp-1})--(\ref{eq:decomp-4}) is completely general and 
independent from U-spin considerations. 
For example, further same-sign contributions in the CF and DCS decays can be absorbed by a 
redefinition of $t_0$ and $t_2$, see Ref.~\cite{Brod:2012ud}.
The meaning as a U-spin expansion only comes into play if we assume a hierarchy for the parameters according to their subscript. 

The letters used to denote the amplitudes should not be
confused with any ideas about the diagrams that generate them. That is,
the use of $p_0$ and $t_0$ is there since in some limit $p_0$ is dominated
by penguin diagrams and $t_0$ by tree diagrams. Yet, this is not always the case, and thus
it is important to keep in mind that all that we do know at this stage
is that the above is a general reparametrization of the decay
amplitudes, and that each amplitude arises at a given order in the
U-spin expansion.
In the topological interpretation of the appearing parameters, $t_0$ includes both tree and exchange 
diagrams, which are absorbed~\cite{Muller:2015lua}. Moreover, $s_1$ contains the broken penguin 
and $p_0$ includes contributions from tree, exchange, penguin and penguin annihilation 
diagrams~\cite{Muller:2015lua,Brod:2012ud}.

We note that 
the U-spin parametrization is completely general when we assume no CPV
in the CF and DCS decays, which is also the case to a very good
approximation in the SM.
Beyond the SM, there can be additional amplitude contributions to 
the $\overline{D}^0\rightarrow K^+\pi^-$ and $\overline{D}^0\rightarrow \pi^+K^-$ decays which come 
with a relative weak phase from CP violating new physics. We do not
discuss this case any further here.

In terms of the above amplitudes, the branching ratios are given as 
\begin{align}
\mathcal{BR}(D\rightarrow P_1P_2 ) &= \vert \mathcal{A}\vert^2 \times
                                     \mathcal{P}(D,P_1,P_2)\,,
                                     \nonumber \\
\mathcal{P}(D,P_1,P_2) &= \tau_D \times \frac{1}{16\pi m_D^3} \sqrt{   
(m_D^2 - (m_{P_1} - m_{P_2})^2) (m_D^2 - ( m_{P_1} + m_{P_2})^2 )
}\,.
\end{align}
The direct CP asymmetries are \cite{Golden:1989qx, Pirtskhalava:2011va, Nierste:2017cua}
\begin{align}
a_{CP}^{\mathrm{dir}} &= \mathrm{Im}\left(\frac{\lambda_b}{\Sigma}\right) \mathrm{Im}\left(\frac{A_b}{A_{\Sigma}}\right)\,.
\end{align}

\section{Solving the complete U-spin System \label{sec:solving} }

We discuss how to extract the U-spin parameters of Eqs.~(\ref{eq:decomp-1})--(\ref{eq:decomp-4}) from the observables.
We are mainly interested in the ratios of parameters and less in their
absolute sizes and 
therefore we consider only quantities normalized on $t_0$, that is
\begin{align} \label{eq:def-u-par}
&\tilde{t}_1 \equiv \frac{t_1}{t_0}\,, \qquad
\tilde{t}_2 \equiv \frac{t_2}{t_0}\,,  \qquad
\tilde{s}_1 \equiv \frac{s_1}{t_0}\,, \qquad 
\tilde{p}_0 \equiv \frac{p_0}{t_0}\,,\qquad
\tilde{p}_1 \equiv \frac{p_1}{t_0}\,.
\end{align}
We choose, without loss of generality, the tree amplitude $t_0$ to be real.
The relative phase between $\mathcal{A}(K\pi)$ and $\mathcal{A}(\pi K)$ is physical and can be extracted in 
experimental measurements. However, the relative phases between $\mathcal{A}(\pi\pi)$, $\mathcal{A}(KK)$ and 
$\mathcal{A}(K\pi)$ are unphysical, i.e. not observable on principal grounds. 
This corresponds to two additional phase choices that can be made in the
U-spin parametrization. 
Consequently, without loss of generality, we can also choose the two parameters $\tilde{s}_1$ and $\tilde{t}_2$ to be real. 
Altogether, that makes eight real parameters, that we want to extract, not counting the normalization $t_0$. 
Of these, four parameters are in the CKM-leading part of the amplitudes and four in the CKM-suppressed one.
In the CP limit $\mathrm{Im}\lambda_b \rightarrow 0$ we can absorb $\tilde{p}_0$ and $\tilde{p}_1$ into 
$\tilde{t}_2$ and $\tilde{s}_1$ respectively, which makes four real parameters in that limit.

The eight parameters can be extracted from eight observables that can be used to completely determine them.
Additional observables can then be used in order to overconstrain the system. 
We divide the eight observables that we use to determine the system into four categories:

$(i)$ Branching ratio measurements (3 observables) \cite{Tanabashi:2018oca}. 
They are used to calculate the squared matrix elements.  We neglect the tiny effects of
order $|\lambda_b/\Sigma|$ and we get 
\begin{align}
\vert A_{\Sigma}(KK)\vert^2 &= 
\frac{\mathcal{B}( \overline{D}^0\rightarrow K^+K^-)) }{ |\Sigma|^2 \mathcal{P}(D^0,K^+,K^-) }\,, \\
\vert A_{\Sigma}( \pi\pi)\vert^2 &= 
\frac{\mathcal{B}( \overline{D}^0\rightarrow \pi^+\pi^-) }{ |\Sigma|^2 \mathcal{P}(D^0,\pi^+,\pi^-) }\,, \\
\vert A( K\pi)\vert^2 &= 
\frac{\mathcal{B}( \overline{D}^0\rightarrow K^+\pi^-) }{ |V_{cs} V_{ud}^*|^2 \mathcal{P}(D^0, K^+, \pi^-) }\,, \\
\vert A(\pi K)\vert^2 &= 
\frac{\mathcal{B}( \overline{D}^0\rightarrow K^-\pi^+) }{ |V_{cd} V_{us}^*|^2 \mathcal{P}(D^0,K^-,\pi^+ )   }\,.
\end{align}
We consider three ratios of combinations of the four branching ratios, which are 
\begin{align}
R_{K\pi} &\equiv \frac{
\vert A(K\pi)\vert^2 - 
\vert A(\pi K)\vert^2
}{ 
\vert A(K\pi) \vert^2 + 
\vert A(\pi K) \vert^2
}\,, \label{eq:br-ratio-1}\\ 
R_{KK,\pi\pi} &\equiv \frac{
\vert A(KK)\vert^2 - 
\vert A(\pi\pi)\vert^2
}{ 
\vert A(KK)\vert^2 + 
\vert A(\pi \pi)\vert^2
}\,,  \label{eq:br-ratio-2} \\
R_{KK,\pi\pi,K\pi} &\equiv \frac{
  \vert A( KK )\vert^2
+ \vert A( \pi\pi)  \vert^2
- \vert A( K\pi)  \vert^2
- \vert A( \pi K) \vert^2
}{
  \vert A( KK) \vert^2
+ \vert A( \pi \pi)  \vert^2
+ \vert A( K \pi)  \vert^2
+ \vert A( \pi K) \vert^2
}\,. \label{eq:br-ratio-3}
\end{align}

$(ii)$ Strong phase which does not require CP violation (1 observable). The relative strong phase between CF and DCS decay modes
\begin{align}
\delta_{K\pi} &\equiv \mathrm{arg}\left( \frac{\mathcal{A}(\overline{D}^0\rightarrow K^-\pi^+)}{\mathcal{A}(D^0\rightarrow K^-\pi^+)}\right)
  = \mathrm{arg}\left(\frac{\mathcal{A}(D^0\rightarrow K^+\pi^-)}{\mathcal{A}(D^0\rightarrow K^-\pi^+)}   \right) 
\end{align}
can be obtained from time-dependent 
measurements~\cite{Chau:1993ec,Browder:1995ay,  Wolfenstein:1995kv, Blaylock:1995ay, Falk:1999ts, Gronau:2000ru, Bergmann:2000id, Falk:2001hx, Grossman:2006jg, Kagan:2009gb, Aaij:2016roz} 
or correlated $D^0 \overline{D}^0$ decays~\cite{Bigi:1986dp, Xing:1996pn, Gronau:2001nr, Atwood:2002ak, Asner:2005wf, Asner:2012xb} 
at a charm-$\tau$ factory.  

$(iii)$ Integrated direct CP asymmetries (2 observables). In
particular we use~\cite{Einhorn:1975fw,
  Abbott:1979fw,Golden:1989qx,  Brod:2012ud, Grinstein:2014aza,
  Franco:2012ck,  Nierste:2017cua, Nierste:2015zra, Muller:2015rna,
  Hiller:2012xm, Grossman:2018ptn, Buccella:1994nf, Grossman:2006jg,
  Artuso:2008vf, Khodjamirian:2017zdu, Cheng:2012wr, Feldmann:2012js,
  Li:2012cfa, Atwood:2012ac, Grossman:2012ry, Buccella:2019kpn,
  Yu:2017oky, Brod:2011re}
\begin{align}
\Delta a_{CP}^{\mathrm{dir}} &\equiv a_{CP}^{\mathrm{dir}}(D^0\rightarrow K^+K^-) - a_{CP}^{\mathrm{dir}}(D^0\rightarrow \pi^+\pi^-)\,, \\
\Sigma a_{CP}^{\mathrm{dir}} &\equiv a_{CP}^{\mathrm{dir}}(D^0\rightarrow K^+K^-) + a_{CP}^{\mathrm{dir}}(D^0\rightarrow \pi^+\pi^-)\,.
\end{align}

$(iv)$ Strong phases that require CP violation (2 observables) 
\cite{Grossman:2006jg, Bergmann:2000id, Kagan:2009gb, Bigi:1986dp, Xing:1996pn, Gronau:2001nr, Atwood:2002ak, Nierste:2015zra}. 
These are the relative phases of the amplitudes of a $\overline{D}^0$ and $D^0$
going into one of the CP eigenstates. They are proportional to CPV
effects and thus very hard to extract. In particular,
\begin{align}
\delta_{KK} &\equiv \mathrm{arg}\left(\frac{\mathcal{A}(\overline{D}^0\rightarrow K^+K^-)}{\mathcal{A}(D^0\rightarrow K^+K^-)} \right)\,, \qquad
\delta_{\pi\pi} \equiv \mathrm{arg}\left(\frac{\mathcal{A}(\overline{D}^0\rightarrow \pi^+\pi^-)}{\mathcal{A}(D^0\rightarrow \pi^+\pi^-)} \right) \,.
\end{align}
These can be obtained from time-dependent measurements or measurements
of correlated $D^0\overline{D}^0$ pairs. 

In principle, using the above observables the system Eqs.~(\ref{eq:decomp-1})--(\ref{eq:decomp-4}) 
is exactly solvable as long as the data is very precise. 
In the CP limit the branching ratio measurements~$(i)$ and the strong phase~$(ii)$ are sufficient to determine
$\tilde{t}_1$, $\tilde{t}_2$ and $\tilde{s}_1$, which are the complete set of independent parameters in this limit. 

For our parameter extraction with current data, we expand the observables to first nonvanishing 
order in the U-spin expansion. We measure the power counting of that
expansion with a generic parameter $\varepsilon$, which, for nominal U-spin
breaking effects is expected to be $\varepsilon \sim 25\%$.
All of the explicit results that we give below have the nice feature that the parameters can be extracted from 
them up to relative corrections of order
$\mathcal{O}(\varepsilon^2)$. Below it is understood that we neglect
all effects of that order.

In terms of our parameters the ratios of branching ratios are given as
\begin{align}
R_{K\pi} &= - \mathrm{Re}(\tilde{t}_1) \,, \\ 
R_{KK,\pi\pi} &= - 2 \tilde{s}_1\,, \\
R_{KK,\pi\pi,K\pi} &=  \frac{1}{2}\left( \tilde{s}_1^2 - \frac{1}{4}\vert \tilde{t}_1\vert^2 +
\tilde{t}_2      \right)\,. 
\end{align}
By inserting the expressions for $R_{K\pi}$ and $R_{KK,\pi\pi}$ into Eq.~(\ref{eq:br-ratio-3}) we can solve the 
above equations for the independent parameter combinations. The result
up to $\mathcal{O}(\varepsilon^2)$ is
\begin{align}
 \mathrm{Re}( \tilde{t}_1 ) &= - R_{K\pi}\,, \label{eq:Ret1tilde}\\
 \tilde{s}_1 &= -\frac{1}{2} R_{KK,\pi\pi}\,, \label{eq:Res1tilde} \\
-\frac{1}{4}  \left(\mathrm{Im}\, \tilde{t}_1\right)^2 
+ \tilde{t}_2  &=  2 R_{KK,\pi\pi,K\pi} - \frac{1}{4}R_{KK,\pi\pi}^2 + \frac{1}{4} R_{K\pi}^2  \,.\label{eq:combi}
\end{align}

We are then able to determine $\tilde{t}_1$ with Eq.~(\ref{eq:Ret1tilde}) and the strong phase between the CF and 
DCS mode, see also Ref.~\cite{Bergmann:2000id},
\begin{align}
\delta_{K\pi} &= \mathrm{arg}\left(- \frac{1-\frac{1}{2} \tilde{t}_1 }{1+\frac{1}{2} \tilde{t}_1 } \right)
       = -\mathrm{Im} (\tilde{t}_1)\,, \label{eq:strongphase} 
\end{align}
where in the last step we neglect terms of relative order of
$\varepsilon^2$.

After that we can  determine $\tilde{s}_1$ and $\tilde{t}_2$ from Eqs.~(\ref{eq:Res1tilde}) and (\ref{eq:combi}), respectively.   
The sum and difference of the integrated direct CP asymmetries can be used together 
with the phases $\delta_{KK}$ and $\delta_{\pi\pi}$ to  determine $\tilde{p}_0$ and $\tilde{p}_1$.
We have 
\begin{align}
\Delta a_{CP}^{\mathrm{dir}} &= \mathrm{Im}\left(\frac{\lambda_b}{\Sigma}\right) \times  
4\,\mathrm{Im}\left(\tilde{p}_0 \right) 
\,, \label{eq:DeltaACPdirParameter}
\end{align}
and
\begin{align}
\Sigma a_{CP}^{\mathrm{dir}} =
  2\, \mathrm{Im}\left(\frac{\lambda_b}{\Sigma}\right) 
\times \left[ 
2\, \mathrm{Im}(\tilde{p}_0 ) \tilde{s}_1 +
     \mathrm{Im}(\tilde{p}_1) \right] \,.
\end{align}
Note that also $\Delta a_{CP}^{\mathrm{dir}}$ and $\Sigma a_{CP}^{\mathrm{dir}}$ share the feature of 
corrections entering only at the relative order $\mathcal{O}(\varepsilon^2)$ compared to the leading result.
The measurement of $\Delta a_{CP}^{\mathrm{dir}}$ is basically a direct measurement of $\mathrm{Im}\,\tilde{p}_0$, 
\begin{align}
\mathrm{Im}\,\tilde{p}_0 &=  \frac{1}{4 \mathrm{Im}(\lambda_b/\Sigma)} \Delta a_{CP}^{\mathrm{dir}}\,. \label{eq:penguinovertree} 
\end{align}

The phases $\delta_{KK}$ and $\delta_{\pi\pi}$ give (see e.g. Ref.~\cite{Nierste:2015zra})
\begin{align}
\mathrm{Re}\left(\frac{A_b(D^0\rightarrow K^+K^-)}{A_{\Sigma}(D^0\rightarrow K^+K^-)} \right) - \mathrm{Re}\left(\frac{A_b(D^0\rightarrow \pi^+\pi^-)}{A_{\Sigma}(D^0\rightarrow \pi^+\pi^-)} \right)  &=
4 \mathrm{Re}(\tilde{p}_0) \,, \label{eq:retildep0}
\end{align}
and
\begin{align}
\mathrm{Re}\left(\frac{A_b(D^0\rightarrow K^+K^-)}{A_{\Sigma}(D^0\rightarrow K^+K^-)} \right) + \mathrm{Re}\left(\frac{A_b(D^0\rightarrow \pi^+\pi^-)}{A_{\Sigma}(D^0\rightarrow \pi^+\pi^-)} \right) &=
2\, \mathrm{Re}(2 \tilde{p}_0 \tilde{s}_1 + \tilde{p}_1 ) \nn\\
&= 2 \left[ 2\, \mathrm{Re}(\tilde{p}_0) \tilde{s}_1 + \mathrm{Re}(\tilde{p}_1)\right] \,. \label{eq:retildep1}
\end{align}
As $\tilde{s}_1$ is already in principle determined from the other observables, this gives us then the full information on $\tilde{p}_0$ and $\tilde{p}_1$.

As the observables $\delta_{KK}$ and $\delta_{\pi\pi}$ are the hardest
to measure, we are not providing here the explicit 
relation of Eq.~(\ref{eq:retildep0}) and Eq.~(\ref{eq:retildep1}) to these observables, acknowledging just that the
corresponding parameter combinations can be determined from these in a straight forward way. 

Taking everything into account, we conclude that the above system of eight
observables for eight parameters can completely be solved. This is done
where the values of the CKM elements are used as inputs.
We emphasize that in principle with correlated double-tag measurements at a future charm-tau factory~\cite{Gronau:2001nr, Goldhaber:1976fp, Bigi:1986dp, Xing:1994mn, Xing:1995vj, Xing:1995vn, Xing:1996pn, Xing:1999yw,  Asner:2005wf, Asner:2008ft, Asner:2012xb, Xing:2019uzz}
we could even overconstrain the system.

\section{Numerical Results \label{sec:numerics} }

We use the formalism introduced in Sec.~\ref{sec:solving} now with the currently available measurements.
As not all of the observables have yet been measured, we cannot
determine all of the U-spin parameters. Yet, we use the ones that we do
have data on to get useful information on some of them.

\begin{itemize}
\item
Using Gaussian error propagation without taking into account correlations, from the branching ratio 
measurements~\cite{Tanabashi:2018oca}
\begin{align}
\mathcal{BR}(D^0\rightarrow K^+K^-)     &= (3.97\pm 0.07) \cdot 10^{-3}\,, \\
\mathcal{BR}(D^0\rightarrow \pi^+\pi^-) &= (1.407\pm 0.025)\cdot 10^{-3}\,, \\
\mathcal{BR}(D^0\rightarrow K^+\pi^- )  &= (1.366 \pm 0.028) \cdot 10^{-4}\,, \\
\mathcal{BR}(D^0\rightarrow K^- \pi^+ )  &= (3.89\pm 0.04)\cdot 10^{-2}\,,
\end{align}
we obtain the normalized combinations
\begin{align}
R_{K\pi}           &= -0.11 \pm 0.01\,,   \\
R_{KK,\pi\pi}      &= 0.534 \pm 0.009\,,  \\ 
R_{KK,\pi\pi,K\pi} &= 0.071 \pm 0.009\,.
\end{align}
\item
The strong phase between DCS and CF mode for the scenario of no CP violation in the DCS mode is \cite{Amhis:2016xyh}
\begin{align}
\delta_{K\pi} &= \left(8.6^{+9.1}_{-9.7}\right) ^{\circ}\,.
\end{align}
\item
The world average of $\Delta a_{CP}^{\mathrm{dir}}$ is given in Eq.~(\ref{eq:HFLAVav}).
\item
The sum of CP asymmetries $\Sigma a_{CP}^{\mathrm{dir}}$ in which CP violation has not yet been observed. 
In order to get an estimate we use the HFLAV averages for the single measurements of the 
CP asymmetries~\cite{Amhis:2016xyh,Aaij:2014gsa,Aaltonen:2011se, Aubert:2007if, Staric:2008rx, Csorna:2001ww, Link:2000aw, Aitala:1997ff} 
\begin{align}
A_{CP}(D^0\rightarrow \pi^+\pi^-) &=  0.0000 \pm 0.0015\,, \\
A_{CP}(D^0\rightarrow K^+K^-)     &= -0.0016 \pm 0.0012\,, 
\end{align}
and subtract the contribution from indirect charm CP violation $a_{CP}^{\mathrm{ind}} = (0.028 \pm 0.026)\%$~\cite{Carbone:2019}.
We obtain 
\begin{align}
\Sigma a_{CP}^{\mathrm{dir}} &= A_{CP}(D^0\rightarrow K^+K^-) +  A_{CP}(D^0\rightarrow \pi^+\pi^-) - 2  a_{CP}^{\mathrm{ind}} \nn\\ 
			     &= -0.002\pm 0.002\,, 
\end{align}
where we do not take into account correlations, which may be sizable.
\item
The phases $\delta_{KK}$ and $\delta_{\pi\pi}$ have not yet been measured, and we cannot get
any indirect information about them.
\end{itemize}

From Eqs.~(\ref{eq:Ret1tilde})--(\ref{eq:strongphase}) it follows that 
\begin{align}
\mathrm{Re}( \tilde{t}_1 ) &= 0.109 \pm 0.011\,, \label{eq:result-ret1tilde}\\
\mathrm{Im}( \tilde{t}_1 ) &= -0.15^{+0.16}_{-0.17}\,, \label{eq:result-imt1tilde} \\ 
\tilde{s}_1 &= -0.2668 \pm 0.0045\,,  \label{eq:result-res1tilde} \\
-\frac{1}{4} \left(\mathrm{Im}\tilde{t}_1\right)^2 + \mathrm{Re}(\tilde{t}_2)  &= 
			0.075\pm  0.018	 \,. \label{eq:result-combi}
\end{align}
Employing~\cite{Tanabashi:2018oca} 
\begin{align}
\mathrm{Im}\left(\frac{\lambda_b}{\Sigma}\right) = (-6.3\pm 0.3)\cdot 10^{-4}\,,
\end{align}
and inserting the measurement of $\Delta a_{CP}^{\mathrm{dir}}$ into
Eq.~(\ref{eq:penguinovertree}), we obtain
\begin{align}
\mathrm{Im}\,\tilde{p}_0 &= 0.65 \pm 0.12 \,. \label{eq:resultp0tilde} 
\end{align}
Using $\Sigma a_{CP}^{\mathrm{dir}} $ we get
\begin{align}
2 \mathrm{Im}(\tilde{p}_0 ) \tilde{s}_1 + \mathrm{Im}(\tilde{p}_1) &=  1.7\pm 1.6\,.   
\end{align}

Few remarks are in order regarding the numerical values we obtained.
\begin{enumerate}
\item
Among the five parameters defined in Eq.~(\ref{eq:def-u-par}),
$\tilde{p}_1$ is the least constrained parameter as we have basically no information
about it.
In order to learn more about it we need measurements of $\Sigma a_{CP}^{\mathrm{dir}}$ 
as well as of the phases $\delta_{KK}$ and $\delta_{\pi\pi}$. 
\item 
The higher order U-spin breaking parameters are
consistently smaller than the first order ones, and the second order ones
are even smaller. This is what we expect assuming the U-spin expansion.
\item 
Eqs.~(\ref{eq:result-ret1tilde})--(\ref{eq:result-combi})
suggest that the SU(3)$_F$ breaking of the tree amplitude $\tilde{t}_1$ is smaller than the broken penguin contained in 
$\tilde{s}_1$. 
\item
Using Eqs.~(\ref{eq:result-ret1tilde})--(\ref{eq:result-combi}) 
we can get a rough estimate for the $\mathcal{O}(\varepsilon^2)$ corrections
that enter the expression for $\Delta a_{CP}^{\mathrm{dir}}$ in Eq.~(\ref{eq:DeltaACPdirParameter}).
The results on the broken penguin suggest that these corrections do not exceed a level of $\sim 10\%$. 
We cannot, however, determine these corrections completely without further knowledge on $\tilde{p}_1$. 
\end{enumerate}

\section{The $\Delta U=0$ rule \label{sec:deltau0rule} }

We now turn to discuss the implications of Eq.~(\ref{eq:resultp0tilde}).
We rewrite
Eq.~(\ref{eq:DeltaACPdirParameter}) as
\begin{align}
\Delta a_{CP}^{\mathrm{dir}} &= 
4\, \mathrm{Im}\left(\frac{\lambda_b}{\Sigma}\right) \left|\tilde{p}_0 \right| \sin( \delta_{\mathrm{strong}})\,,
\end{align}
with the unknown strong phase 
\begin{align}
\delta_{\mathrm{strong}} &= \mathrm{arg}(\tilde{p}_0)\,.
\end{align}
Then the numerical result in Eq.~(\ref{eq:resultp0tilde}) reads 
\begin{align}
\left|\tilde{p}_0\right| \sin( \delta_{\mathrm{strong}}) &= 0.65 \pm 0.12\,. \label{eq:mainresult}
\end{align}
Recall that in the group theoretical language the parameters $t_0$ and $p_0$ are 
the matrix elements of the $\Delta U=1$ and $\Delta U=0$ operators, respectively~\cite{Brod:2011re}.
For the ratio of the matrix elements of these operators we employ now the following parametrization
\begin{equation} \label{eq:defC}
\tilde p_0 = B + C e^{i \delta}\,,
\end{equation}
such that $B$ is the short-distance (SD) ratio and the second term arises
from long-distance (LD) effects. While the separation between SD and LD
is not well-defined, what we
have in mind here is that diagrams with a $b$ quark in the loop are perturbative
and those with quarks lighter than the charm are not.

In Eq.~(\ref{eq:DeltaI12-generic}) of Sec.~\ref{sec:DeltaI12inKDB} below we apply the same decomposition into a \lq\lq{}no QCD\rq\rq{} part and corrections to that also to the $\Delta I=1/2$ rules in $K$, $D$ and $B$ decays to pions. It is instructive to compare all of these systems in the same language.

We first argue that in Eq.~(\ref{eq:defC}) to a very good approximation $B=1$. This is
basically the statement that perturbatively, the diagrams with
intermediate $b$ are tiny. More explicitly, in that case, that is when
we neglect the SD $b$ penguins, we have 
\begin{align}
Q^{\Delta U=1} \equiv \frac{Q^{\bar{s}s} - Q^{\bar{d}d}}{2}\,,\qquad 
Q^{\Delta U=0} \equiv \frac{Q^{\bar{s}s} + Q^{\bar{d}d}}{2}\,. 
\end{align}
Setting $C=0$ then corresponds to the statement that
only $Q^{\bar{s}s}$ can produce $K^+K^-$ and only $Q^{\bar{d}d}$ 
can produce $\pi^+\pi^-$. This implies that for $C=0$
\begin{align}
\bra{K^+K^-} Q^{\bar{d}d} \ket{D^0} &=  
\bra{\pi^+\pi^-} Q^{\bar{s}s} \ket{D^0} = 0\,, \label{eq:assumption2-1} 
\end{align}
and 
\begin{align}
\bra{K^+K^-} Q^{\bar{s}s} \ket{D^0} \neq 0\,, \qquad
\bra{\pi^+\pi^-} Q^{\bar{d}d} \ket{D^0} \neq 0\,. \label{eq:assumption2-2}
\end{align}
We then see that $B=1$ since
\begin{align}
\frac{\bra{K^+K^-} Q^{\Delta U=0}\ket{D^0}
}{
\bra{K^+K^-} Q^{\Delta U=1}\ket{D^0}
}  &= 1\,,  \qquad
\frac{  \bra{\pi^+\pi^-} Q^{\Delta U=0}\ket{D^0}
}{
\bra{\pi^+\pi^-} Q^{\Delta U=1}\ket{D^0}
}  = -1\,. \label{eq:quarkmodel-wo-phase}
\end{align}
We note that in the SU(3)$_F$ limit we also have
\begin{align}
\bra{K^+K^-} Q^{\Delta U=1}\ket{D^0} &= - \bra{\pi^+\pi^-} Q^{\Delta U=1}\ket{D^0}\,, \\
\bra{K^+K^-} Q^{\Delta U=0}\ket{D^0} &= \bra{\pi^+\pi^-} Q^{\Delta U=0}\ket{D^0}\,,
\end{align}
but this is not used to argue that $B=1$.

We then argue that $\delta \sim \mathcal{O}(1)$. The reason is that
non-perturbative effects involve on-shell particles, or in other words,
rescattering, and such effects give rise to large strong phases to the
LD effects
independent of the magnitude of the LD amplitude.

In the case that $B=1$, $\delta \sim \mathcal{O}(1)$ and using the fact that the CKM ratios are
small we conclude that the CP asymmetry is roughly given by the CKM
factor times $C$
\begin{align}
\Delta a_{CP}^{\mathrm{dir}} = 4\,
  \mathrm{Im}\left(\frac{\lambda_b}{\Sigma}\right) \times C \times
  \sin \delta\,. 
\end{align}
Now the question is: what is $C$? 
As at this time no method is available in order to calculate $C$ with a well-defined theoretical uncertainty, 
we do not employ here a dynamical calculation in order to provide a SM prediction for $C$ and $\Delta a_{CP}^{\mathrm{dir}}$.
We rather show the different principal possibilities and how to interpret them in view of the current data.
In order to do so we measure the order of magnitude of the QCD correction term $C$ relative to the \lq\lq{}no QCD\rq\rq{} limit 
$\tilde{p}_0=1$. Relative to that limit, we differentiate between three cases
\begin{enumerate}
\item $C = \mathcal{O}(\alpha_s/\pi)$: Perturbative corrections  to
  $\tilde p_0$.
\item $C = \mathcal{O}(1)$: Non-perturbative corrections that produce
  strong phases from rescattering but do not significantly change the
  magnitude of $\tilde p_0$.
\item $C \gg \mathcal{O}(1)$: Large non-perturbative effects with
  significant magnitude changes and strong phases from rescattering to
  $\tilde p_0$.
\end{enumerate}
Note that category (2) and (3) are in principle not different, as they both include non-perturbative effects, which 
differ only in their size.

Some perturbative results concluded that $C=\mathcal{O}(\alpha_s/\pi)$, leading to 
$\Delta a_{CP}^{\mathrm{dir}}\sim 10^{-4}$ \cite{Grossman:2006jg, Bigi:2011re}.
Note that the value  $\Delta a_{CP}^{\mathrm{dir}} = 1\times 10^{-4}$,
assuming $O(1)$ strong phase, 
would correspond numerically to
$C \sim 0.04$. We conclude that if there is a good argument that $C$ is of category~(1), the measurement of $\Delta a_{CP}^{\mathrm{dir}}$ would be a sign of beyond the SM (BSM) physics, because it would indicate a relative $\mathcal{O}(10)$ enhancement. 

If the value of $\Delta a_{CP}^{\mathrm{dir}}$ would have turned out as large as 
suggested by the central value of some (statistically unsignificant) earlier measurements~\cite{Aaij:2011in,Collaboration:2012qw},
we would clearly need category~(3) in order to explain that, i.e. penguin diagrams that are 
enhanced in magnitude, see e.g. Refs.~\cite{Brod:2012ud, Hiller:2012xm, Cheng:2012wr, Feldmann:2012js, Li:2012cfa, Atwood:2012ac, Grossman:2012ry, Brod:2011re}. Another example for category (3) is the $\Delta I=1/2$ rule in the kaon sector 
which is further discussed in sections \ref{sec:DeltaI12inKDB} and \ref{sec:UandIrules}. 

The current data, Eq.~(\ref{eq:mainresult}), is 
consistent with category (2).
In the SM picture, the measurement of $\Delta a_{CP}^{\mathrm{dir}}$ proves the non-perturbative nature of the $\Delta U=0$ matrix 
elements with a mild enhancement from $\mathcal{O}(1)$ rescattering effects. This is the $\Delta U=0$ rule for charm. 

Note that the predictions for $\Delta a_{CP}^{\mathrm{dir}}$ of category (i) and (ii) differ by $\mathcal{O}(10)$, although 
category (ii) contains only an $\mathcal{O}(1)$ nonperturbative enhancement with respect to the \lq\lq{}no QCD\rq\rq{} limit 
$\tilde{p}_0=1$. We emphasize that a measure for a QCD enhancement is not necessarily its impact on an observable, but the amplitude level comparison with the absence of QCD effects. 

We also mention that we do not need SU(3)$_F$ breaking effects to
explain the data.
Yet, the observation of 
$\vert \tilde{s}_1\vert > \vert \tilde{t}_1\vert$ in 
Eqs.~(\ref{eq:result-ret1tilde})--(\ref{eq:result-res1tilde})
provide additional supporting evidence that rescattering is
significant. 
Though no proof of the $\Delta U=0$ rule on its own, 
this matches its upshot and is indicative of the importance of rescattering effects also 
in the broken penguin which is contained in $\tilde{s}_1$.

With future data on the phases $\delta_{KK}$ and $\delta_{\pi\pi}$ we will be 
able to determine the strong phase~$\delta$ of Eq.~(\ref{eq:defC}). In that way it will be possible to completely determine 
the characteristics of the emerging $\Delta U=0$ rule.

\section{$\Delta I=1/2$ Rules in $K$, $D$ and $B$ Decays \label{sec:DeltaI12inKDB}}

It is instructive to compare the $\Delta U=0$ rule in charm with the
$\Delta I=1/2$ rule in kaon physics, and furthermore also to the corresponding ratios of isospin matrix elements 
of $D$ and $B$ decays.
For a review of the $\Delta I=1/2$ rule see e.g. Ref.~\cite{Buras:2014maa}.

In kaon physics we consider $K \to\pi\pi$ decays. Employing an isospin parametrization we have~\cite{Buras:2014maa}
\begin{align}  
\A(K^+\rightarrow \pi^+\pi^0) &= \frac{3}{2} A_2^K e^{i\delta_2^K}\,, \nonumber\\
\A(K^0\rightarrow \pi^+\pi^-) &= A_0^K e^{i\delta_0^K} + \sqrt{\frac{1}{2}} A_2^K e^{i\delta_2^K}\,, \nonumber \\
\A(K^0\rightarrow \pi^0\pi^0) &= A_0^K e^{i\delta_0^K} - \sqrt{2} A_2^K e^{i\delta_2^K}\,. \label{eq:kaondata}
\end{align}
Note that the strong phases of $A_0^K$ and $A_2^K$ are factored out, so that $A_{0,2}^K$ contain weak phases only.
The data give
\begin{align} 
\left|\frac{A_0^K}{A_2^K}\right| \approx 22.35\,,\qquad
\delta_0^K - \delta_2^K = (47.5\pm 0.9)^{\circ}\,, \label{eq:kaon-deltaI12-rule}
\end{align}
see Ref.~\cite{Buras:2014maa} and references therein for more
details. $A_{0,2}^K$ have a small imaginary part stemming from the CKM matrix elements only. 
To a very good approximation the real parts $\mathrm{Re}(A_0^K)$ and $\mathrm{Re}(A_2^K)$ in the $\Delta I=1/2$ rule 
depend only on the tree operators~\cite{Buras:2015yba, Kitahara:2016nld}
\begin{align}
Q_1 &= (\bar{s}_{\alpha} u_{\beta})_{V-A} (\bar{u}_{\beta} d_{\alpha})_{V-A}\,, \qquad 
Q_2  = (\bar{s} u)_{V-A} (\bar{u} d)_{V-A}\,. 
\end{align}
The lattice results Refs.~\cite{Bai:2015nea, Blum:2015ywa,Boyle:2012ys} show an emerging physical interpretation of the $\Delta I=1/2$ rule, that is an approximate cancellation of two contributions in $\mathrm{Re}(A_2^K)$, which does not
take place in $\mathrm{Re}(A_0^K)$. These two contributions are different color contractions of the same operator.

The isospin decompositions of $D\rightarrow \pi\pi$ and $B\rightarrow \pi\pi$ are completely analog to Eq.~(\ref{eq:kaondata}).
To differentiate the charm and beauty isospin decompositions from the kaon one, we put the corresponding superscripts
to the respective analog matrix elements. Leaving away the superscripts indicates generic formulas that are valid for all three 
meson systems. 

In order to understand better the anatomy of the $\Delta I=1/2$ rule we use again the form
\begin{align}
\frac{A_0}{A_2} &= B + C e^{i\delta}\,, \label{eq:DeltaI12-generic}
\end{align}
analogously to Eq.~(\ref{eq:defC}) in Sec.~\ref{sec:deltau0rule} for the $\Delta U=0$ rule. 
Here, $B$ is again the contribution in the limit of \lq\lq{}no QCD\rq\rq{}, and $C e^{i\delta}$ contains the 
corrections to that limit.
Now, as discussed in Refs.~\cite{Buras:1988ky,Buras:2014maa}, in the limit of no strong interactions
only the $Q_2$ operator contributes in Eq.~(\ref{eq:DeltaI12-generic}). 
Note that the operator $Q_1$ is only generated from QCD corrections.
When we switch off QCD, the amplitude into neutral pions vanishes and we have for $K,D,B\rightarrow \pi\pi$ 
equally~\cite{Buras:1988ky,Buras:2014maa}
\begin{align}
B &= \sqrt{2}\,. \label{eq:kaon-deltaI12-rule-no-qcd}
\end{align}
This corresponds to the limit $\tilde{p}_0 = 1$ that we considered in Sec.~\ref{sec:deltau0rule} for the $\Delta U=0$ rule.
The exact numerical value in Eq.~(\ref{eq:kaon-deltaI12-rule-no-qcd}) of course depends on the convention used for 
the normalization of $A_{0,2}$ in the isospin decomposition Eq.~(\ref{eq:kaondata}), where 
we use the one present in the literature.

For the isospin decomposition of $D^+\rightarrow \pi^+\pi^0$, $D^0\rightarrow \pi^+\pi^-$ and 
$D^0\rightarrow \pi^0\pi^0$, we simply combine the fit of Ref.~\cite{Franco:2012ck} to get 
\begin{align}
\left|\frac{A_0^D}{A_2^D}\right| &= 2.47\pm 0.07\,, \qquad  
\delta_0^D - \delta_2^D = (\pm 93 \pm 3)^{\circ}\,.  \label{eq:charm-deltaI12-rule}
\end{align}
Reproducing the $\Delta I=1/2$ rule for charm Eq.~(\ref{eq:charm-deltaI12-rule}) is an optimal future testing ground for 
emerging new interesting non-perturbative methods~\cite{Khodjamirian:2017zdu}. 
Very promising steps on a conceptual level are also taken by lattice QCD~\cite{Hansen:2012tf}.

In $K$ and $D$ decays the contributions of penguin operators to $A_0$ is CKM-suppressed, i.e. to a good 
approximation $A_0$ is generated from tree operators only.
In $B$ decays the situation is more involved because there is no relative hierarchy between the relevant CKM matrix elements.
However, one can separate tree and penguin contributions by including the measurements of CP asymmetries within a global fit,
as done in Ref.~\cite{Grinstein:2014aza}. From Fig.~3 therein we find for the ratio of matrix elements of tree operators that 
\begin{align}
\left| \frac{A_0^B}{A_2^B}\right| \sim \sqrt{2}
\end{align} 
is well compatible with the data, the best fit point having $\vert A_0^B / A_2^B\vert = 1.5$.
The fit result for the phase difference $\delta_0^B - \delta_2^B$ is not given in Ref.~\cite{Grinstein:2014aza}.

The emerging picture is: 
The $\Delta I=1/2$ rule in $B$ decays is compatible or close to the \lq\lq{}no QCD\rq\rq{} limit.
The $\Delta I=1/2$ rule in kaon physics clearly belongs to category~(3) of Sec.~\ref{sec:deltau0rule}.
Here, the non-perturbative rescattering affects not only the phases but also the magnitudes of the corresponding matrix elements.
Finally, the $\Delta I=1/2$ rule in charm decays is intermediate and shows an $\mathcal{O}(1)$ enhancement, 
similar to the $\Delta U=0$ rule that we found in Sec.~\ref{sec:deltau0rule}.

We can understand these differences from the different mass scales that govern $K$, $D$ and $B$ decays. 
Rescattering effects are most important in $K$ decays, less important but still significant in $D$ decays, 
and small in $B$ decays.

\section{Phenomenology of the $\Delta U=0$ vs. $\Delta I=1/2$ rule \label{sec:UandIrules}}

An interesting difference between the $\Delta I=1/2$ rule in kaon decays and the 
$\Delta U=0$ rule in charm decays is their effect on the phenomenology. 
Large rescattering enhances the CP violation
effects in $D$ decays, but it reduces the effect in kaon decays. 
The reason for the difference lies in the fact that in kaon decays the
SD decay generates only a $u \bar u$ final state, while in charm decays
it generates to a very good approximation the same amount of $d \bar d$
and $s \bar s$ states.

We write the amplitudes very generally and up to a normalization factor as
\begin{equation}
{\cal A} = 1 + r a e^{i(\phi+\delta)}\,, \label{eq:generic-ampl}
\end{equation}
such that $r$ is real and depends on CKM matrix elements, $a$ is real
and corresponds to the ratio of the respective
hadronic matrix elements, $\phi$ is a weak phase and
$\delta$ is a strong phase. For kaons $a$ is the ratio of matrix elements of the operators 
$Q^{\Delta I=1/2}$ over $Q^{\Delta I=3/2}$, while for charm it is the ratio of matrix elements of the operators  
$Q^{\Delta U=0}$ over $Q^{\Delta U=1}$. 

We first consider the case where we neglect the third
generation. In that limit
for kaons we have the decomposition
\begin{equation}
{\cal A}_K = V_{us}V_{ud}^* (A_{1/2} + r_{CG} A_{3/2})\,,
\end{equation}
where $r_{CG}$ is the CG coefficient that can be read from Eq.~(\ref{eq:kaondata}).
For charm we have
\begin{equation}
{\cal A}_D = V_{cs}V_{us}^* A_1.
\end{equation}
That means that in the two-generational limit for kaons we have $r=1$ 
and in charm $r=0$. If we switch on the third generation we get small corrections to these
values in each case: $r\ll 1$ for charm and $|r-1|\ll 1$ for
kaons. These effects come from the non-unitarity of the $2 \times 2$
CKM. For the kaon case   
there is an extra effect that stems from SD penguins that
come with $V_{ts}V_{td}^*$. In both cases we have $\delta \sim
\mathcal{O}(1)$ from non-perturbative rescattering, as
well as $\phi \sim \mathcal{O}(1)$.

The general formula for direct CP asymmetry is given as~\cite{Tanabashi:2018oca}
\begin{align}
A_{CP} &= -\frac{2 r a  \sin(\delta) \sin(\phi)
}{ 
1 + (ra)^2+ 2 ra \cos(\delta) \cos(\phi)
}
\approx
\begin{cases}
2 r a  \sin(\delta) \sin(\phi)&{\mbox{for $ra \ll 1$}},\\
2 (r a)^{-1}  \sin(\delta) \sin(\phi)&{\mbox{for $ra \gg 1$}}.\\
\end{cases} \label{eq:CPasym-general-formula} 
\end{align}
Non-perturbative effects enhance $a$ in both kaon and charm decays.
This means the effect which is visible in the CP asymmetry is different depending on the value of $r$.
For $r a\ll 1$ increasing $a$ results in enhancement of the CP
asymmetry, while for $r a \gg 1$ it is suppressed. 
These two cases correspond to the charm and kaon cases,
respectively. It follows that the $\Delta I=1/2$ rule in kaons reduces
CP violating effects, while the $\Delta U=0$ rule in charm enhances them.

\section{Conclusions \label{sec:conclusions}}

From the recent determination of $\Delta a_{CP}^{\mathrm{dir}}$ we derive the ratio of $\Delta U=0$ over $\Delta U=1$ 
amplitudes as  
\begin{align}
\vert \tilde{p}_0\vert \sin(\delta_{\mathrm{strong}}) &= 0.65 \pm 0.12\,.\label{eq:mainresult-conclusion}
\end{align}
In principle two options are possible in order to explain this result: In the perturbative 
picture beyond the SM (BSM) physics is necessary to explain Eq.~(\ref{eq:mainresult-conclusion}). 
On the other hand, in the SM picture, we find that all that is
required in order to explain the result is a 
mild non-perturbative enhancement due to rescattering
effects. Therefore, it is hard to argue that BSM physics is required. 

Our interpretation of the result is that the measurement of $\Delta a_{CP}^{\mathrm{dir}}$ 
provides a proof for the $\Delta U=0$ rule in charm.
The enhancement of the  $\Delta U=0$ amplitude is not as significant
as the one present in the $\Delta I=1/2$ rule for kaons. 
In the future, with more information on the strong phase of $\tilde{p}_0$ from time-dependent measurements 
or measurements of correlated $D^0\overline{D}^0$ decays, we will be able to completely
determine the extent of the $\Delta U=0$ rule.

Interpreting the result within the SM implies that we expect a moderate
non-perturbative effect and nominal $SU(3)_F$ breaking. The former fact
implies that we expect U-spin invariant strong phases to be
$\mathcal{O}(1)$. The latter implies that we anticipate the yet to be
determined  $SU(3)_F$ breaking effects not to be large. Thus, there are
two qualitative predictions we can make
\begin{align}
\delta_{\mathrm{strong}}\sim
\mathcal{O}(1), \qquad
a_{CP}^{\mathrm{dir}} (D^0\rightarrow K^+K^-) \approx -a_{CP}^{\mathrm{dir}}(D^0\rightarrow \pi^+\pi^-)\,.
\end{align}
Verifying these predictions will make the SM interpretation of the
data more solid.

\begin{acknowledgments}
We thank Alex Kagan, Yossi Nir, Giovanni Punzi and Alan Schwartz for useful discussions.
The work of YG is supported in part by the NSF grant PHY1316222.
SS is supported by a DFG Forschungs\-stipendium under contract no. SCHA 2125/1-1.
\end{acknowledgments}

\bibliography{uspin-DeltaACP.bib}

\begin{thebibliography}{79}%
\makeatletter
\providecommand \@ifxundefined [1]{%
 \@ifx{#1\undefined}
}%
\providecommand \@ifnum [1]{%
 \ifnum #1\expandafter \@firstoftwo
 \else \expandafter \@secondoftwo
 \fi
}%
\providecommand \@ifx [1]{%
 \ifx #1\expandafter \@firstoftwo
 \else \expandafter \@secondoftwo
 \fi
}%
\providecommand \natexlab [1]{#1}%
\providecommand \enquote  [1]{``#1''}%
\providecommand \bibnamefont  [1]{#1}%
\providecommand \bibfnamefont [1]{#1}%
\providecommand \citenamefont [1]{#1}%
\providecommand \href@noop [0]{\@secondoftwo}%
\providecommand \href [0]{\begingroup \@sanitize@url \@href}%
\providecommand \@href[1]{\@@startlink{#1}\@@href}%
\providecommand \@@href[1]{\endgroup#1\@@endlink}%
\providecommand \@sanitize@url [0]{\catcode `\\12\catcode `\$12\catcode
  `\&12\catcode `\#12\catcode `\^12\catcode `\_12\catcode `\%12\relax}%
\providecommand \@@startlink[1]{}%
\providecommand \@@endlink[0]{}%
\providecommand \url  [0]{\begingroup\@sanitize@url \@url }%
\providecommand \@url [1]{\endgroup\@href {#1}{\urlprefix }}%
\providecommand \urlprefix  [0]{URL }%
\providecommand \Eprint [0]{\href }%
\providecommand \doibase [0]{http://dx.doi.org/}%
\providecommand \selectlanguage [0]{\@gobble}%
\providecommand \bibinfo  [0]{\@secondoftwo}%
\providecommand \bibfield  [0]{\@secondoftwo}%
\providecommand \translation [1]{[#1]}%
\providecommand \BibitemOpen [0]{}%
\providecommand \bibitemStop [0]{}%
\providecommand \bibitemNoStop [0]{.\EOS\space}%
\providecommand \EOS [0]{\spacefactor3000\relax}%
\providecommand \BibitemShut  [1]{\csname bibitem#1\endcsname}%
\let\auto@bib@innerbib\@empty
\bibitem [{\citenamefont {Aaij}\ \emph {et~al.}(2019)\citenamefont {Aaij} \emph
  {et~al.}}]{Aaij:2019kcg}%
  \BibitemOpen
  \bibfield  {author} {\bibinfo {author} {\bibfnamefont {R.}~\bibnamefont
  {Aaij}} \emph {et~al.} (\bibinfo {collaboration} {LHCb}),\ }\href@noop {} {\
  (\bibinfo {year} {2019})},\ \Eprint {http://arxiv.org/abs/1903.08726}
  {arXiv:1903.08726 [hep-ex]} \BibitemShut {NoStop}%
\bibitem [{\citenamefont {Aitala}\ \emph {et~al.}(1998)\citenamefont {Aitala}
  \emph {et~al.}}]{Aitala:1997ff}%
  \BibitemOpen
  \bibfield  {author} {\bibinfo {author} {\bibfnamefont {E.~M.}\ \bibnamefont
  {Aitala}} \emph {et~al.} (\bibinfo {collaboration} {E791}),\ }\href {\doibase
  10.1016/S0370-2693(97)01570-0} {\bibfield  {journal} {\bibinfo  {journal}
  {Phys. Lett.}\ }\textbf {\bibinfo {volume} {B421}},\ \bibinfo {pages} {405}
  (\bibinfo {year} {1998})},\ \Eprint {http://arxiv.org/abs/hep-ex/9711003}
  {arXiv:hep-ex/9711003 [hep-ex]} \BibitemShut {NoStop}%
\bibitem [{\citenamefont {Link}\ \emph {et~al.}(2000)\citenamefont {Link} \emph
  {et~al.}}]{Link:2000aw}%
  \BibitemOpen
  \bibfield  {author} {\bibinfo {author} {\bibfnamefont {J.~M.}\ \bibnamefont
  {Link}} \emph {et~al.} (\bibinfo {collaboration} {FOCUS}),\ }\href {\doibase
  10.1016/S0370-2693(00)01237-5, 10.1016/S0370-2693(00)01039-X} {\bibfield
  {journal} {\bibinfo  {journal} {Phys. Lett.}\ }\textbf {\bibinfo {volume}
  {B491}},\ \bibinfo {pages} {232} (\bibinfo {year} {2000})},\ \bibinfo {note}
  {[Erratum: Phys. Lett.B495,443(2000)]},\ \Eprint
  {http://arxiv.org/abs/hep-ex/0005037} {arXiv:hep-ex/0005037 [hep-ex]}
  \BibitemShut {NoStop}%
\bibitem [{\citenamefont {Csorna}\ \emph {et~al.}(2002)\citenamefont {Csorna}
  \emph {et~al.}}]{Csorna:2001ww}%
  \BibitemOpen
  \bibfield  {author} {\bibinfo {author} {\bibfnamefont {S.~E.}\ \bibnamefont
  {Csorna}} \emph {et~al.} (\bibinfo {collaboration} {CLEO}),\ }\href {\doibase
  10.1103/PhysRevD.65.092001} {\bibfield  {journal} {\bibinfo  {journal} {Phys.
  Rev.}\ }\textbf {\bibinfo {volume} {D65}},\ \bibinfo {pages} {092001}
  (\bibinfo {year} {2002})},\ \Eprint {http://arxiv.org/abs/hep-ex/0111024}
  {arXiv:hep-ex/0111024 [hep-ex]} \BibitemShut {NoStop}%
\bibitem [{\citenamefont {Aubert}\ \emph {et~al.}(2008)\citenamefont {Aubert}
  \emph {et~al.}}]{Aubert:2007if}%
  \BibitemOpen
  \bibfield  {author} {\bibinfo {author} {\bibfnamefont {B.}~\bibnamefont
  {Aubert}} \emph {et~al.} (\bibinfo {collaboration} {BaBar}),\ }\href
  {\doibase 10.1103/PhysRevLett.100.061803} {\bibfield  {journal} {\bibinfo
  {journal} {Phys. Rev. Lett.}\ }\textbf {\bibinfo {volume} {100}},\ \bibinfo
  {pages} {061803} (\bibinfo {year} {2008})},\ \Eprint
  {http://arxiv.org/abs/0709.2715} {arXiv:0709.2715 [hep-ex]} \BibitemShut
  {NoStop}%
\bibitem [{\citenamefont {Staric}\ \emph {et~al.}(2008)\citenamefont {Staric}
  \emph {et~al.}}]{Staric:2008rx}%
  \BibitemOpen
  \bibfield  {author} {\bibinfo {author} {\bibfnamefont {M.}~\bibnamefont
  {Staric}} \emph {et~al.} (\bibinfo {collaboration} {Belle}),\ }\href
  {\doibase 10.1016/j.physletb.2008.10.052} {\bibfield  {journal} {\bibinfo
  {journal} {Phys. Lett.}\ }\textbf {\bibinfo {volume} {B670}},\ \bibinfo
  {pages} {190} (\bibinfo {year} {2008})},\ \Eprint
  {http://arxiv.org/abs/0807.0148} {arXiv:0807.0148 [hep-ex]} \BibitemShut
  {NoStop}%
\bibitem [{\citenamefont {Aaltonen}\ \emph
  {et~al.}(2012{\natexlab{a}})\citenamefont {Aaltonen} \emph
  {et~al.}}]{Aaltonen:2011se}%
  \BibitemOpen
  \bibfield  {author} {\bibinfo {author} {\bibfnamefont {T.}~\bibnamefont
  {Aaltonen}} \emph {et~al.} (\bibinfo {collaboration} {CDF}),\ }\href
  {\doibase 10.1103/PhysRevD.85.012009} {\bibfield  {journal} {\bibinfo
  {journal} {Phys. Rev.}\ }\textbf {\bibinfo {volume} {D85}},\ \bibinfo {pages}
  {012009} (\bibinfo {year} {2012}{\natexlab{a}})},\ \Eprint
  {http://arxiv.org/abs/1111.5023} {arXiv:1111.5023 [hep-ex]} \BibitemShut
  {NoStop}%
\bibitem [{\citenamefont {Aaltonen}\ \emph
  {et~al.}(2012{\natexlab{b}})\citenamefont {Aaltonen} \emph
  {et~al.}}]{Collaboration:2012qw}%
  \BibitemOpen
  \bibfield  {author} {\bibinfo {author} {\bibfnamefont {T.}~\bibnamefont
  {Aaltonen}} \emph {et~al.} (\bibinfo {collaboration} {CDF}),\ }\href
  {\doibase 10.1103/PhysRevLett.109.111801} {\bibfield  {journal} {\bibinfo
  {journal} {Phys. Rev. Lett.}\ }\textbf {\bibinfo {volume} {109}},\ \bibinfo
  {pages} {111801} (\bibinfo {year} {2012}{\natexlab{b}})},\ \Eprint
  {http://arxiv.org/abs/1207.2158} {arXiv:1207.2158 [hep-ex]} \BibitemShut
  {NoStop}%
\bibitem [{\citenamefont {Aaij}\ \emph {et~al.}(2012)\citenamefont {Aaij} \emph
  {et~al.}}]{Aaij:2011in}%
  \BibitemOpen
  \bibfield  {author} {\bibinfo {author} {\bibfnamefont {R.}~\bibnamefont
  {Aaij}} \emph {et~al.} (\bibinfo {collaboration} {LHCb}),\ }\href {\doibase
  10.1103/PhysRevLett.108.129903, 10.1103/PhysRevLett.108.111602} {\bibfield
  {journal} {\bibinfo  {journal} {Phys. Rev. Lett.}\ }\textbf {\bibinfo
  {volume} {108}},\ \bibinfo {pages} {111602} (\bibinfo {year} {2012})},\
  \Eprint {http://arxiv.org/abs/1112.0938} {arXiv:1112.0938 [hep-ex]}
  \BibitemShut {NoStop}%
\bibitem [{\citenamefont {Aaij}\ \emph {et~al.}(2013)\citenamefont {Aaij} \emph
  {et~al.}}]{Aaij:2013bra}%
  \BibitemOpen
  \bibfield  {author} {\bibinfo {author} {\bibfnamefont {R.}~\bibnamefont
  {Aaij}} \emph {et~al.} (\bibinfo {collaboration} {LHCb}),\ }\href {\doibase
  10.1016/j.physletb.2013.04.061} {\bibfield  {journal} {\bibinfo  {journal}
  {Phys. Lett.}\ }\textbf {\bibinfo {volume} {B723}},\ \bibinfo {pages} {33}
  (\bibinfo {year} {2013})},\ \Eprint {http://arxiv.org/abs/1303.2614}
  {arXiv:1303.2614 [hep-ex]} \BibitemShut {NoStop}%
\bibitem [{\citenamefont {Aaij}\ \emph {et~al.}(2014)\citenamefont {Aaij} \emph
  {et~al.}}]{Aaij:2014gsa}%
  \BibitemOpen
  \bibfield  {author} {\bibinfo {author} {\bibfnamefont {R.}~\bibnamefont
  {Aaij}} \emph {et~al.} (\bibinfo {collaboration} {LHCb}),\ }\href {\doibase
  10.1007/JHEP07(2014)041} {\bibfield  {journal} {\bibinfo  {journal} {JHEP}\
  }\textbf {\bibinfo {volume} {07}},\ \bibinfo {pages} {041} (\bibinfo {year}
  {2014})},\ \Eprint {http://arxiv.org/abs/1405.2797} {arXiv:1405.2797
  [hep-ex]} \BibitemShut {NoStop}%
\bibitem [{\citenamefont {Aaij}\ \emph {et~al.}(2016)\citenamefont {Aaij} \emph
  {et~al.}}]{Aaij:2016cfh}%
  \BibitemOpen
  \bibfield  {author} {\bibinfo {author} {\bibfnamefont {R.}~\bibnamefont
  {Aaij}} \emph {et~al.} (\bibinfo {collaboration} {LHCb}),\ }\href {\doibase
  10.1103/PhysRevLett.116.191601} {\bibfield  {journal} {\bibinfo  {journal}
  {Phys. Rev. Lett.}\ }\textbf {\bibinfo {volume} {116}},\ \bibinfo {pages}
  {191601} (\bibinfo {year} {2016})},\ \Eprint
  {http://arxiv.org/abs/1602.03160} {arXiv:1602.03160 [hep-ex]} \BibitemShut
  {NoStop}%
\bibitem [{\citenamefont {Aaij}\ \emph
  {et~al.}(2017{\natexlab{a}})\citenamefont {Aaij} \emph
  {et~al.}}]{Aaij:2016dfb}%
  \BibitemOpen
  \bibfield  {author} {\bibinfo {author} {\bibfnamefont {R.}~\bibnamefont
  {Aaij}} \emph {et~al.} (\bibinfo {collaboration} {LHCb}),\ }\href {\doibase
  10.1016/j.physletb.2017.01.061} {\bibfield  {journal} {\bibinfo  {journal}
  {Phys. Lett.}\ }\textbf {\bibinfo {volume} {B767}},\ \bibinfo {pages} {177}
  (\bibinfo {year} {2017}{\natexlab{a}})},\ \Eprint
  {http://arxiv.org/abs/1610.09476} {arXiv:1610.09476 [hep-ex]} \BibitemShut
  {NoStop}%
\bibitem [{\citenamefont {Amhis}\ \emph {et~al.}(2017)\citenamefont {Amhis}
  \emph {et~al.}}]{Amhis:2016xyh}%
  \BibitemOpen
  \bibfield  {author} {\bibinfo {author} {\bibfnamefont {Y.}~\bibnamefont
  {Amhis}} \emph {et~al.} (\bibinfo {collaboration} {HFLAV}),\ }\href {\doibase
  10.1140/epjc/s10052-017-5058-4} {\bibfield  {journal} {\bibinfo  {journal}
  {Eur. Phys. J.}\ }\textbf {\bibinfo {volume} {C77}},\ \bibinfo {pages} {895}
  (\bibinfo {year} {2017})},\ \Eprint {http://arxiv.org/abs/1612.07233}
  {arXiv:1612.07233 [hep-ex]} \BibitemShut {NoStop}%
\bibitem [{\citenamefont {{A. Carbone}}(2019)}]{Carbone:2019}%
  \BibitemOpen
  \bibfield  {author} {\bibinfo {author} {\bibnamefont {{A. Carbone}}},\
  }\href@noop {} {\enquote {\bibinfo {title} {{Recent LHCb results on CP
  violation}},}\ } (\bibinfo {year} {2019}),\ \bibinfo {note} {{LHC Seminar, 21
  Mar 2019, CERN, Switzerland}}\BibitemShut {NoStop}%
\bibitem [{\citenamefont {Tanabashi}\ \emph {et~al.}(2018)\citenamefont
  {Tanabashi} \emph {et~al.}}]{Tanabashi:2018oca}%
  \BibitemOpen
  \bibfield  {author} {\bibinfo {author} {\bibfnamefont {M.}~\bibnamefont
  {Tanabashi}} \emph {et~al.} (\bibinfo {collaboration} {Particle Data
  Group}),\ }\href {\doibase 10.1103/PhysRevD.98.030001} {\bibfield  {journal}
  {\bibinfo  {journal} {Phys. Rev.}\ }\textbf {\bibinfo {volume} {D98}},\
  \bibinfo {pages} {030001} (\bibinfo {year} {2018})}\BibitemShut {NoStop}%
\bibitem [{\citenamefont {Gell-Mann}\ and\ \citenamefont
  {Pais}(1955)}]{GellMann:1955jx}%
  \BibitemOpen
  \bibfield  {author} {\bibinfo {author} {\bibfnamefont {M.}~\bibnamefont
  {Gell-Mann}}\ and\ \bibinfo {author} {\bibfnamefont {A.}~\bibnamefont
  {Pais}},\ }\href {\doibase 10.1103/PhysRev.97.1387} {\bibfield  {journal}
  {\bibinfo  {journal} {Phys. Rev.}\ }\textbf {\bibinfo {volume} {97}},\
  \bibinfo {pages} {1387} (\bibinfo {year} {1955})}\BibitemShut {NoStop}%
\bibitem [{\citenamefont {Gell-Mann}\ and\ \citenamefont
  {Rosenfeld}(1957)}]{GellMann:1957wh}%
  \BibitemOpen
  \bibfield  {author} {\bibinfo {author} {\bibfnamefont {M.}~\bibnamefont
  {Gell-Mann}}\ and\ \bibinfo {author} {\bibfnamefont {A.~H.}\ \bibnamefont
  {Rosenfeld}},\ }\href {\doibase 10.1146/annurev.ns.07.120157.002203}
  {\bibfield  {journal} {\bibinfo  {journal} {Ann. Rev. Nucl. Part. Sci.}\
  }\textbf {\bibinfo {volume} {7}},\ \bibinfo {pages} {407} (\bibinfo {year}
  {1957})}\BibitemShut {NoStop}%
\bibitem [{\citenamefont {Gaillard}\ and\ \citenamefont
  {Lee}(1974)}]{Gaillard:1974nj}%
  \BibitemOpen
  \bibfield  {author} {\bibinfo {author} {\bibfnamefont {M.~K.}\ \bibnamefont
  {Gaillard}}\ and\ \bibinfo {author} {\bibfnamefont {B.~W.}\ \bibnamefont
  {Lee}},\ }\href {\doibase 10.1103/PhysRevLett.33.108} {\bibfield  {journal}
  {\bibinfo  {journal} {Phys. Rev. Lett.}\ }\textbf {\bibinfo {volume} {33}},\
  \bibinfo {pages} {108} (\bibinfo {year} {1974})}\BibitemShut {NoStop}%
\bibitem [{\citenamefont {Bardeen}\ \emph {et~al.}(1987)\citenamefont
  {Bardeen}, \citenamefont {Buras},\ and\ \citenamefont
  {Gerard}}]{Bardeen:1986vz}%
  \BibitemOpen
  \bibfield  {author} {\bibinfo {author} {\bibfnamefont {W.~A.}\ \bibnamefont
  {Bardeen}}, \bibinfo {author} {\bibfnamefont {A.~J.}\ \bibnamefont {Buras}},
  \ and\ \bibinfo {author} {\bibfnamefont {J.~M.}\ \bibnamefont {Gerard}},\
  }\href {\doibase 10.1016/0370-2693(87)91156-7} {\bibfield  {journal}
  {\bibinfo  {journal} {Phys. Lett.}\ }\textbf {\bibinfo {volume} {B192}},\
  \bibinfo {pages} {138} (\bibinfo {year} {1987})}\BibitemShut {NoStop}%
\bibitem [{\citenamefont {Buras}\ \emph {et~al.}(2014)\citenamefont {Buras},
  \citenamefont {Gerard},\ and\ \citenamefont {Bardeen}}]{Buras:2014maa}%
  \BibitemOpen
  \bibfield  {author} {\bibinfo {author} {\bibfnamefont {A.~J.}\ \bibnamefont
  {Buras}}, \bibinfo {author} {\bibfnamefont {J.-M.}\ \bibnamefont {Gerard}}, \
  and\ \bibinfo {author} {\bibfnamefont {W.~A.}\ \bibnamefont {Bardeen}},\
  }\href {\doibase 10.1140/epjc/s10052-014-2871-x} {\bibfield  {journal}
  {\bibinfo  {journal} {Eur. Phys. J.}\ }\textbf {\bibinfo {volume} {C74}},\
  \bibinfo {pages} {2871} (\bibinfo {year} {2014})},\ \Eprint
  {http://arxiv.org/abs/1401.1385} {arXiv:1401.1385 [hep-ph]} \BibitemShut
  {NoStop}%
\bibitem [{\citenamefont {Bai}\ \emph {et~al.}(2015)\citenamefont {Bai} \emph
  {et~al.}}]{Bai:2015nea}%
  \BibitemOpen
  \bibfield  {author} {\bibinfo {author} {\bibfnamefont {Z.}~\bibnamefont
  {Bai}} \emph {et~al.} (\bibinfo {collaboration} {RBC, UKQCD}),\ }\href
  {\doibase 10.1103/PhysRevLett.115.212001} {\bibfield  {journal} {\bibinfo
  {journal} {Phys. Rev. Lett.}\ }\textbf {\bibinfo {volume} {115}},\ \bibinfo
  {pages} {212001} (\bibinfo {year} {2015})},\ \Eprint
  {http://arxiv.org/abs/1505.07863} {arXiv:1505.07863 [hep-lat]} \BibitemShut
  {NoStop}%
\bibitem [{\citenamefont {Blum}\ \emph {et~al.}(2015)\citenamefont {Blum} \emph
  {et~al.}}]{Blum:2015ywa}%
  \BibitemOpen
  \bibfield  {author} {\bibinfo {author} {\bibfnamefont {T.}~\bibnamefont
  {Blum}} \emph {et~al.},\ }\href {\doibase 10.1103/PhysRevD.91.074502}
  {\bibfield  {journal} {\bibinfo  {journal} {Phys. Rev.}\ }\textbf {\bibinfo
  {volume} {D91}},\ \bibinfo {pages} {074502} (\bibinfo {year} {2015})},\
  \Eprint {http://arxiv.org/abs/1502.00263} {arXiv:1502.00263 [hep-lat]}
  \BibitemShut {NoStop}%
\bibitem [{\citenamefont {Boyle}\ \emph {et~al.}(2013)\citenamefont {Boyle}
  \emph {et~al.}}]{Boyle:2012ys}%
  \BibitemOpen
  \bibfield  {author} {\bibinfo {author} {\bibfnamefont {P.~A.}\ \bibnamefont
  {Boyle}} \emph {et~al.} (\bibinfo {collaboration} {RBC, UKQCD}),\ }\href
  {\doibase 10.1103/PhysRevLett.110.152001} {\bibfield  {journal} {\bibinfo
  {journal} {Phys. Rev. Lett.}\ }\textbf {\bibinfo {volume} {110}},\ \bibinfo
  {pages} {152001} (\bibinfo {year} {2013})},\ \Eprint
  {http://arxiv.org/abs/1212.1474} {arXiv:1212.1474 [hep-lat]} \BibitemShut
  {NoStop}%
\bibitem [{\citenamefont {Buras}\ \emph {et~al.}(2015)\citenamefont {Buras},
  \citenamefont {Gorbahn}, \citenamefont {{J\"ager}},\ and\ \citenamefont
  {Jamin}}]{Buras:2015yba}%
  \BibitemOpen
  \bibfield  {author} {\bibinfo {author} {\bibfnamefont {A.~J.}\ \bibnamefont
  {Buras}}, \bibinfo {author} {\bibfnamefont {M.}~\bibnamefont {Gorbahn}},
  \bibinfo {author} {\bibfnamefont {S.}~\bibnamefont {{J\"ager}}}, \ and\
  \bibinfo {author} {\bibfnamefont {M.}~\bibnamefont {Jamin}},\ }\href
  {\doibase 10.1007/JHEP11(2015)202} {\bibfield  {journal} {\bibinfo  {journal}
  {JHEP}\ }\textbf {\bibinfo {volume} {11}},\ \bibinfo {pages} {202} (\bibinfo
  {year} {2015})},\ \Eprint {http://arxiv.org/abs/1507.06345} {arXiv:1507.06345
  [hep-ph]} \BibitemShut {NoStop}%
\bibitem [{\citenamefont {Kitahara}\ \emph {et~al.}(2016)\citenamefont
  {Kitahara}, \citenamefont {Nierste},\ and\ \citenamefont
  {Tremper}}]{Kitahara:2016nld}%
  \BibitemOpen
  \bibfield  {author} {\bibinfo {author} {\bibfnamefont {T.}~\bibnamefont
  {Kitahara}}, \bibinfo {author} {\bibfnamefont {U.}~\bibnamefont {Nierste}}, \
  and\ \bibinfo {author} {\bibfnamefont {P.}~\bibnamefont {Tremper}},\ }\href
  {\doibase 10.1007/JHEP12(2016)078} {\bibfield  {journal} {\bibinfo  {journal}
  {JHEP}\ }\textbf {\bibinfo {volume} {12}},\ \bibinfo {pages} {078} (\bibinfo
  {year} {2016})},\ \Eprint {http://arxiv.org/abs/1607.06727} {arXiv:1607.06727
  [hep-ph]} \BibitemShut {NoStop}%
\bibitem [{\citenamefont {Einhorn}\ and\ \citenamefont
  {Quigg}(1975)}]{Einhorn:1975fw}%
  \BibitemOpen
  \bibfield  {author} {\bibinfo {author} {\bibfnamefont {M.~B.}\ \bibnamefont
  {Einhorn}}\ and\ \bibinfo {author} {\bibfnamefont {C.}~\bibnamefont
  {Quigg}},\ }\href {\doibase 10.1103/PhysRevD.12.2015} {\bibfield  {journal}
  {\bibinfo  {journal} {Phys. Rev.}\ }\textbf {\bibinfo {volume} {D12}},\
  \bibinfo {pages} {2015} (\bibinfo {year} {1975})}\BibitemShut {NoStop}%
\bibitem [{\citenamefont {Abbott}\ \emph {et~al.}(1980)\citenamefont {Abbott},
  \citenamefont {Sikivie},\ and\ \citenamefont {Wise}}]{Abbott:1979fw}%
  \BibitemOpen
  \bibfield  {author} {\bibinfo {author} {\bibfnamefont {L.~F.}\ \bibnamefont
  {Abbott}}, \bibinfo {author} {\bibfnamefont {P.}~\bibnamefont {Sikivie}}, \
  and\ \bibinfo {author} {\bibfnamefont {M.~B.}\ \bibnamefont {Wise}},\ }\href
  {\doibase 10.1103/PhysRevD.21.768} {\bibfield  {journal} {\bibinfo  {journal}
  {Phys. Rev.}\ }\textbf {\bibinfo {volume} {D21}},\ \bibinfo {pages} {768}
  (\bibinfo {year} {1980})}\BibitemShut {NoStop}%
\bibitem [{\citenamefont {Golden}\ and\ \citenamefont
  {Grinstein}(1989)}]{Golden:1989qx}%
  \BibitemOpen
  \bibfield  {author} {\bibinfo {author} {\bibfnamefont {M.}~\bibnamefont
  {Golden}}\ and\ \bibinfo {author} {\bibfnamefont {B.}~\bibnamefont
  {Grinstein}},\ }\href {\doibase 10.1016/0370-2693(89)90353-5} {\bibfield
  {journal} {\bibinfo  {journal} {Phys. Lett.}\ }\textbf {\bibinfo {volume}
  {B222}},\ \bibinfo {pages} {501} (\bibinfo {year} {1989})}\BibitemShut
  {NoStop}%
\bibitem [{\citenamefont {Brod}\ \emph
  {et~al.}(2012{\natexlab{a}})\citenamefont {Brod}, \citenamefont {Grossman},
  \citenamefont {Kagan},\ and\ \citenamefont {Zupan}}]{Brod:2012ud}%
  \BibitemOpen
  \bibfield  {author} {\bibinfo {author} {\bibfnamefont {J.}~\bibnamefont
  {Brod}}, \bibinfo {author} {\bibfnamefont {Y.}~\bibnamefont {Grossman}},
  \bibinfo {author} {\bibfnamefont {A.~L.}\ \bibnamefont {Kagan}}, \ and\
  \bibinfo {author} {\bibfnamefont {J.}~\bibnamefont {Zupan}},\ }\href
  {\doibase 10.1007/JHEP10(2012)161} {\bibfield  {journal} {\bibinfo  {journal}
  {JHEP}\ }\textbf {\bibinfo {volume} {10}},\ \bibinfo {pages} {161} (\bibinfo
  {year} {2012}{\natexlab{a}})},\ \Eprint {http://arxiv.org/abs/1203.6659}
  {arXiv:1203.6659 [hep-ph]} \BibitemShut {NoStop}%
\bibitem [{\citenamefont {Grinstein}\ \emph {et~al.}(2014)\citenamefont
  {Grinstein}, \citenamefont {Pirtskhalava}, \citenamefont {Stone},\ and\
  \citenamefont {Uttayarat}}]{Grinstein:2014aza}%
  \BibitemOpen
  \bibfield  {author} {\bibinfo {author} {\bibfnamefont {B.}~\bibnamefont
  {Grinstein}}, \bibinfo {author} {\bibfnamefont {D.}~\bibnamefont
  {Pirtskhalava}}, \bibinfo {author} {\bibfnamefont {D.}~\bibnamefont {Stone}},
  \ and\ \bibinfo {author} {\bibfnamefont {P.}~\bibnamefont {Uttayarat}},\
  }\href {\doibase 10.1103/PhysRevD.89.114014} {\bibfield  {journal} {\bibinfo
  {journal} {Phys. Rev.}\ }\textbf {\bibinfo {volume} {D89}},\ \bibinfo {pages}
  {114014} (\bibinfo {year} {2014})},\ \Eprint {http://arxiv.org/abs/1402.1164}
  {arXiv:1402.1164 [hep-ph]} \BibitemShut {NoStop}%
\bibitem [{\citenamefont {Bhattacharya}\ \emph {et~al.}(2012)\citenamefont
  {Bhattacharya}, \citenamefont {Gronau},\ and\ \citenamefont
  {Rosner}}]{Bhattacharya:2012ah}%
  \BibitemOpen
  \bibfield  {author} {\bibinfo {author} {\bibfnamefont {B.}~\bibnamefont
  {Bhattacharya}}, \bibinfo {author} {\bibfnamefont {M.}~\bibnamefont
  {Gronau}}, \ and\ \bibinfo {author} {\bibfnamefont {J.~L.}\ \bibnamefont
  {Rosner}},\ }\href {\doibase 10.1103/PhysRevD.85.079901,
  10.1103/PhysRevD.85.054014} {\bibfield  {journal} {\bibinfo  {journal} {Phys.
  Rev.}\ }\textbf {\bibinfo {volume} {D85}},\ \bibinfo {pages} {054014}
  (\bibinfo {year} {2012})},\ \bibinfo {note} {[Phys.
  Rev.D85,no.7,079901(2012)]},\ \Eprint {http://arxiv.org/abs/1201.2351}
  {arXiv:1201.2351 [hep-ph]} \BibitemShut {NoStop}%
\bibitem [{\citenamefont {Franco}\ \emph {et~al.}(2012)\citenamefont {Franco},
  \citenamefont {Mishima},\ and\ \citenamefont {Silvestrini}}]{Franco:2012ck}%
  \BibitemOpen
  \bibfield  {author} {\bibinfo {author} {\bibfnamefont {E.}~\bibnamefont
  {Franco}}, \bibinfo {author} {\bibfnamefont {S.}~\bibnamefont {Mishima}}, \
  and\ \bibinfo {author} {\bibfnamefont {L.}~\bibnamefont {Silvestrini}},\
  }\href {\doibase 10.1007/JHEP05(2012)140} {\bibfield  {journal} {\bibinfo
  {journal} {JHEP}\ }\textbf {\bibinfo {volume} {05}},\ \bibinfo {pages} {140}
  (\bibinfo {year} {2012})},\ \Eprint {http://arxiv.org/abs/1203.3131}
  {arXiv:1203.3131 [hep-ph]} \BibitemShut {NoStop}%
\bibitem [{\citenamefont {Hiller}\ \emph {et~al.}(2013)\citenamefont {Hiller},
  \citenamefont {Jung},\ and\ \citenamefont {Schacht}}]{Hiller:2012xm}%
  \BibitemOpen
  \bibfield  {author} {\bibinfo {author} {\bibfnamefont {G.}~\bibnamefont
  {Hiller}}, \bibinfo {author} {\bibfnamefont {M.}~\bibnamefont {Jung}}, \ and\
  \bibinfo {author} {\bibfnamefont {S.}~\bibnamefont {Schacht}},\ }\href
  {\doibase 10.1103/PhysRevD.87.014024} {\bibfield  {journal} {\bibinfo
  {journal} {Phys. Rev.}\ }\textbf {\bibinfo {volume} {D87}},\ \bibinfo {pages}
  {014024} (\bibinfo {year} {2013})},\ \Eprint {http://arxiv.org/abs/1211.3734}
  {arXiv:1211.3734 [hep-ph]} \BibitemShut {NoStop}%
\bibitem [{\citenamefont {Nierste}\ and\ \citenamefont
  {Schacht}(2017)}]{Nierste:2017cua}%
  \BibitemOpen
  \bibfield  {author} {\bibinfo {author} {\bibfnamefont {U.}~\bibnamefont
  {Nierste}}\ and\ \bibinfo {author} {\bibfnamefont {S.}~\bibnamefont
  {Schacht}},\ }\href {\doibase 10.1103/PhysRevLett.119.251801} {\bibfield
  {journal} {\bibinfo  {journal} {Phys. Rev. Lett.}\ }\textbf {\bibinfo
  {volume} {119}},\ \bibinfo {pages} {251801} (\bibinfo {year} {2017})},\
  \Eprint {http://arxiv.org/abs/1708.03572} {arXiv:1708.03572 [hep-ph]}
  \BibitemShut {NoStop}%
\bibitem [{\citenamefont {Nierste}\ and\ \citenamefont
  {Schacht}(2015)}]{Nierste:2015zra}%
  \BibitemOpen
  \bibfield  {author} {\bibinfo {author} {\bibfnamefont {U.}~\bibnamefont
  {Nierste}}\ and\ \bibinfo {author} {\bibfnamefont {S.}~\bibnamefont
  {Schacht}},\ }\href {\doibase 10.1103/PhysRevD.92.054036} {\bibfield
  {journal} {\bibinfo  {journal} {Phys. Rev.}\ }\textbf {\bibinfo {volume}
  {D92}},\ \bibinfo {pages} {054036} (\bibinfo {year} {2015})},\ \Eprint
  {http://arxiv.org/abs/1508.00074} {arXiv:1508.00074 [hep-ph]} \BibitemShut
  {NoStop}%
\bibitem [{\citenamefont {{M\"uller}}\ \emph
  {et~al.}(2015{\natexlab{a}})\citenamefont {{M\"uller}}, \citenamefont
  {Nierste},\ and\ \citenamefont {Schacht}}]{Muller:2015rna}%
  \BibitemOpen
  \bibfield  {author} {\bibinfo {author} {\bibfnamefont {S.}~\bibnamefont
  {{M\"uller}}}, \bibinfo {author} {\bibfnamefont {U.}~\bibnamefont {Nierste}},
  \ and\ \bibinfo {author} {\bibfnamefont {S.}~\bibnamefont {Schacht}},\ }\href
  {\doibase 10.1103/PhysRevLett.115.251802} {\bibfield  {journal} {\bibinfo
  {journal} {Phys. Rev. Lett.}\ }\textbf {\bibinfo {volume} {115}},\ \bibinfo
  {pages} {251802} (\bibinfo {year} {2015}{\natexlab{a}})},\ \Eprint
  {http://arxiv.org/abs/1506.04121} {arXiv:1506.04121 [hep-ph]} \BibitemShut
  {NoStop}%
\bibitem [{\citenamefont {Grossman}\ and\ \citenamefont
  {Schacht}(2019)}]{Grossman:2018ptn}%
  \BibitemOpen
  \bibfield  {author} {\bibinfo {author} {\bibfnamefont {Y.}~\bibnamefont
  {Grossman}}\ and\ \bibinfo {author} {\bibfnamefont {S.}~\bibnamefont
  {Schacht}},\ }\href {\doibase 10.1103/PhysRevD.99.033005} {\bibfield
  {journal} {\bibinfo  {journal} {Phys. Rev.}\ }\textbf {\bibinfo {volume}
  {D99}},\ \bibinfo {pages} {033005} (\bibinfo {year} {2019})},\ \Eprint
  {http://arxiv.org/abs/1811.11188} {arXiv:1811.11188 [hep-ph]} \BibitemShut
  {NoStop}%
\bibitem [{\citenamefont {Buccella}\ \emph {et~al.}(1995)\citenamefont
  {Buccella}, \citenamefont {Lusignoli}, \citenamefont {Miele}, \citenamefont
  {Pugliese},\ and\ \citenamefont {Santorelli}}]{Buccella:1994nf}%
  \BibitemOpen
  \bibfield  {author} {\bibinfo {author} {\bibfnamefont {F.}~\bibnamefont
  {Buccella}}, \bibinfo {author} {\bibfnamefont {M.}~\bibnamefont {Lusignoli}},
  \bibinfo {author} {\bibfnamefont {G.}~\bibnamefont {Miele}}, \bibinfo
  {author} {\bibfnamefont {A.}~\bibnamefont {Pugliese}}, \ and\ \bibinfo
  {author} {\bibfnamefont {P.}~\bibnamefont {Santorelli}},\ }\href {\doibase
  10.1103/PhysRevD.51.3478} {\bibfield  {journal} {\bibinfo  {journal} {Phys.
  Rev.}\ }\textbf {\bibinfo {volume} {D51}},\ \bibinfo {pages} {3478} (\bibinfo
  {year} {1995})},\ \Eprint {http://arxiv.org/abs/hep-ph/9411286}
  {arXiv:hep-ph/9411286 [hep-ph]} \BibitemShut {NoStop}%
\bibitem [{\citenamefont {Grossman}\ \emph {et~al.}(2007)\citenamefont
  {Grossman}, \citenamefont {Kagan},\ and\ \citenamefont
  {Nir}}]{Grossman:2006jg}%
  \BibitemOpen
  \bibfield  {author} {\bibinfo {author} {\bibfnamefont {Y.}~\bibnamefont
  {Grossman}}, \bibinfo {author} {\bibfnamefont {A.~L.}\ \bibnamefont {Kagan}},
  \ and\ \bibinfo {author} {\bibfnamefont {Y.}~\bibnamefont {Nir}},\ }\href
  {\doibase 10.1103/PhysRevD.75.036008} {\bibfield  {journal} {\bibinfo
  {journal} {Phys. Rev.}\ }\textbf {\bibinfo {volume} {D75}},\ \bibinfo {pages}
  {036008} (\bibinfo {year} {2007})},\ \Eprint
  {http://arxiv.org/abs/hep-ph/0609178} {arXiv:hep-ph/0609178 [hep-ph]}
  \BibitemShut {NoStop}%
\bibitem [{\citenamefont {Artuso}\ \emph {et~al.}(2008)\citenamefont {Artuso},
  \citenamefont {Meadows},\ and\ \citenamefont {Petrov}}]{Artuso:2008vf}%
  \BibitemOpen
  \bibfield  {author} {\bibinfo {author} {\bibfnamefont {M.}~\bibnamefont
  {Artuso}}, \bibinfo {author} {\bibfnamefont {B.}~\bibnamefont {Meadows}}, \
  and\ \bibinfo {author} {\bibfnamefont {A.~A.}\ \bibnamefont {Petrov}},\
  }\href {\doibase 10.1146/annurev.nucl.58.110707.171131} {\bibfield  {journal}
  {\bibinfo  {journal} {Ann. Rev. Nucl. Part. Sci.}\ }\textbf {\bibinfo
  {volume} {58}},\ \bibinfo {pages} {249} (\bibinfo {year} {2008})},\ \Eprint
  {http://arxiv.org/abs/0802.2934} {arXiv:0802.2934 [hep-ph]} \BibitemShut
  {NoStop}%
\bibitem [{\citenamefont {Khodjamirian}\ and\ \citenamefont
  {Petrov}(2017)}]{Khodjamirian:2017zdu}%
  \BibitemOpen
  \bibfield  {author} {\bibinfo {author} {\bibfnamefont {A.}~\bibnamefont
  {Khodjamirian}}\ and\ \bibinfo {author} {\bibfnamefont {A.~A.}\ \bibnamefont
  {Petrov}},\ }\href {\doibase 10.1016/j.physletb.2017.09.070} {\bibfield
  {journal} {\bibinfo  {journal} {Phys. Lett.}\ }\textbf {\bibinfo {volume}
  {B774}},\ \bibinfo {pages} {235} (\bibinfo {year} {2017})},\ \Eprint
  {http://arxiv.org/abs/1706.07780} {arXiv:1706.07780 [hep-ph]} \BibitemShut
  {NoStop}%
\bibitem [{\citenamefont {Buccella}\ \emph {et~al.}(2013)\citenamefont
  {Buccella}, \citenamefont {Lusignoli}, \citenamefont {Pugliese},\ and\
  \citenamefont {Santorelli}}]{Buccella:2013tya}%
  \BibitemOpen
  \bibfield  {author} {\bibinfo {author} {\bibfnamefont {F.}~\bibnamefont
  {Buccella}}, \bibinfo {author} {\bibfnamefont {M.}~\bibnamefont {Lusignoli}},
  \bibinfo {author} {\bibfnamefont {A.}~\bibnamefont {Pugliese}}, \ and\
  \bibinfo {author} {\bibfnamefont {P.}~\bibnamefont {Santorelli}},\ }\href
  {\doibase 10.1103/PhysRevD.88.074011} {\bibfield  {journal} {\bibinfo
  {journal} {Phys. Rev.}\ }\textbf {\bibinfo {volume} {D88}},\ \bibinfo {pages}
  {074011} (\bibinfo {year} {2013})},\ \Eprint {http://arxiv.org/abs/1305.7343}
  {arXiv:1305.7343 [hep-ph]} \BibitemShut {NoStop}%
\bibitem [{\citenamefont {Cheng}\ and\ \citenamefont
  {Chiang}(2012)}]{Cheng:2012wr}%
  \BibitemOpen
  \bibfield  {author} {\bibinfo {author} {\bibfnamefont {H.-Y.}\ \bibnamefont
  {Cheng}}\ and\ \bibinfo {author} {\bibfnamefont {C.-W.}\ \bibnamefont
  {Chiang}},\ }\href {\doibase 10.1103/PhysRevD.85.079903,
  10.1103/PhysRevD.85.034036} {\bibfield  {journal} {\bibinfo  {journal} {Phys.
  Rev.}\ }\textbf {\bibinfo {volume} {D85}},\ \bibinfo {pages} {034036}
  (\bibinfo {year} {2012})},\ \bibinfo {note} {[Erratum: Phys.
  Rev.D85,079903(2012)]},\ \Eprint {http://arxiv.org/abs/1201.0785}
  {arXiv:1201.0785 [hep-ph]} \BibitemShut {NoStop}%
\bibitem [{\citenamefont {Feldmann}\ \emph {et~al.}(2012)\citenamefont
  {Feldmann}, \citenamefont {Nandi},\ and\ \citenamefont
  {Soni}}]{Feldmann:2012js}%
  \BibitemOpen
  \bibfield  {author} {\bibinfo {author} {\bibfnamefont {T.}~\bibnamefont
  {Feldmann}}, \bibinfo {author} {\bibfnamefont {S.}~\bibnamefont {Nandi}}, \
  and\ \bibinfo {author} {\bibfnamefont {A.}~\bibnamefont {Soni}},\ }\href
  {\doibase 10.1007/JHEP06(2012)007} {\bibfield  {journal} {\bibinfo  {journal}
  {JHEP}\ }\textbf {\bibinfo {volume} {06}},\ \bibinfo {pages} {007} (\bibinfo
  {year} {2012})},\ \Eprint {http://arxiv.org/abs/1202.3795} {arXiv:1202.3795
  [hep-ph]} \BibitemShut {NoStop}%
\bibitem [{\citenamefont {Li}\ \emph {et~al.}(2012)\citenamefont {Li},
  \citenamefont {Lu},\ and\ \citenamefont {Yu}}]{Li:2012cfa}%
  \BibitemOpen
  \bibfield  {author} {\bibinfo {author} {\bibfnamefont {H.-n.}\ \bibnamefont
  {Li}}, \bibinfo {author} {\bibfnamefont {C.-D.}\ \bibnamefont {Lu}}, \ and\
  \bibinfo {author} {\bibfnamefont {F.-S.}\ \bibnamefont {Yu}},\ }\href
  {\doibase 10.1103/PhysRevD.86.036012} {\bibfield  {journal} {\bibinfo
  {journal} {Phys. Rev.}\ }\textbf {\bibinfo {volume} {D86}},\ \bibinfo {pages}
  {036012} (\bibinfo {year} {2012})},\ \Eprint {http://arxiv.org/abs/1203.3120}
  {arXiv:1203.3120 [hep-ph]} \BibitemShut {NoStop}%
\bibitem [{\citenamefont {Atwood}\ and\ \citenamefont
  {Soni}(2013)}]{Atwood:2012ac}%
  \BibitemOpen
  \bibfield  {author} {\bibinfo {author} {\bibfnamefont {D.}~\bibnamefont
  {Atwood}}\ and\ \bibinfo {author} {\bibfnamefont {A.}~\bibnamefont {Soni}},\
  }\href {\doibase 10.1093/ptep/ptt065} {\bibfield  {journal} {\bibinfo
  {journal} {PTEP}\ }\textbf {\bibinfo {volume} {2013}},\ \bibinfo {pages}
  {093B05} (\bibinfo {year} {2013})},\ \Eprint {http://arxiv.org/abs/1211.1026}
  {arXiv:1211.1026 [hep-ph]} \BibitemShut {NoStop}%
\bibitem [{\citenamefont {Grossman}\ and\ \citenamefont
  {Robinson}(2013)}]{Grossman:2012ry}%
  \BibitemOpen
  \bibfield  {author} {\bibinfo {author} {\bibfnamefont {Y.}~\bibnamefont
  {Grossman}}\ and\ \bibinfo {author} {\bibfnamefont {D.~J.}\ \bibnamefont
  {Robinson}},\ }\href {\doibase 10.1007/JHEP04(2013)067} {\bibfield  {journal}
  {\bibinfo  {journal} {JHEP}\ }\textbf {\bibinfo {volume} {04}},\ \bibinfo
  {pages} {067} (\bibinfo {year} {2013})},\ \Eprint
  {http://arxiv.org/abs/1211.3361} {arXiv:1211.3361 [hep-ph]} \BibitemShut
  {NoStop}%
\bibitem [{\citenamefont {Buccella}\ \emph {et~al.}(2019)\citenamefont
  {Buccella}, \citenamefont {Paul},\ and\ \citenamefont
  {Santorelli}}]{Buccella:2019kpn}%
  \BibitemOpen
  \bibfield  {author} {\bibinfo {author} {\bibfnamefont {F.}~\bibnamefont
  {Buccella}}, \bibinfo {author} {\bibfnamefont {A.}~\bibnamefont {Paul}}, \
  and\ \bibinfo {author} {\bibfnamefont {P.}~\bibnamefont {Santorelli}},\
  }\href@noop {} {\  (\bibinfo {year} {2019})},\ \Eprint
  {http://arxiv.org/abs/1902.05564} {arXiv:1902.05564 [hep-ph]} \BibitemShut
  {NoStop}%
\bibitem [{\citenamefont {Yu}\ \emph {et~al.}(2017)\citenamefont {Yu},
  \citenamefont {Wang},\ and\ \citenamefont {Li}}]{Yu:2017oky}%
  \BibitemOpen
  \bibfield  {author} {\bibinfo {author} {\bibfnamefont {F.-S.}\ \bibnamefont
  {Yu}}, \bibinfo {author} {\bibfnamefont {D.}~\bibnamefont {Wang}}, \ and\
  \bibinfo {author} {\bibfnamefont {H.-n.}\ \bibnamefont {Li}},\ }\href
  {\doibase 10.1103/PhysRevLett.119.181802} {\bibfield  {journal} {\bibinfo
  {journal} {Phys. Rev. Lett.}\ }\textbf {\bibinfo {volume} {119}},\ \bibinfo
  {pages} {181802} (\bibinfo {year} {2017})},\ \Eprint
  {http://arxiv.org/abs/1707.09297} {arXiv:1707.09297 [hep-ph]} \BibitemShut
  {NoStop}%
\bibitem [{\citenamefont {Brod}\ \emph
  {et~al.}(2012{\natexlab{b}})\citenamefont {Brod}, \citenamefont {Kagan},\
  and\ \citenamefont {Zupan}}]{Brod:2011re}%
  \BibitemOpen
  \bibfield  {author} {\bibinfo {author} {\bibfnamefont {J.}~\bibnamefont
  {Brod}}, \bibinfo {author} {\bibfnamefont {A.~L.}\ \bibnamefont {Kagan}}, \
  and\ \bibinfo {author} {\bibfnamefont {J.}~\bibnamefont {Zupan}},\ }\href
  {\doibase 10.1103/PhysRevD.86.014023} {\bibfield  {journal} {\bibinfo
  {journal} {Phys. Rev.}\ }\textbf {\bibinfo {volume} {D86}},\ \bibinfo {pages}
  {014023} (\bibinfo {year} {2012}{\natexlab{b}})},\ \Eprint
  {http://arxiv.org/abs/1111.5000} {arXiv:1111.5000 [hep-ph]} \BibitemShut
  {NoStop}%
\bibitem [{\citenamefont {{M\"uller}}\ \emph
  {et~al.}(2015{\natexlab{b}})\citenamefont {{M\"uller}}, \citenamefont
  {Nierste},\ and\ \citenamefont {Schacht}}]{Muller:2015lua}%
  \BibitemOpen
  \bibfield  {author} {\bibinfo {author} {\bibfnamefont {S.}~\bibnamefont
  {{M\"uller}}}, \bibinfo {author} {\bibfnamefont {U.}~\bibnamefont {Nierste}},
  \ and\ \bibinfo {author} {\bibfnamefont {S.}~\bibnamefont {Schacht}},\ }\href
  {\doibase 10.1103/PhysRevD.92.014004} {\bibfield  {journal} {\bibinfo
  {journal} {Phys. Rev.}\ }\textbf {\bibinfo {volume} {D92}},\ \bibinfo {pages}
  {014004} (\bibinfo {year} {2015}{\natexlab{b}})},\ \Eprint
  {http://arxiv.org/abs/1503.06759} {arXiv:1503.06759 [hep-ph]} \BibitemShut
  {NoStop}%
\bibitem [{\citenamefont {Pirtskhalava}\ and\ \citenamefont
  {Uttayarat}(2012)}]{Pirtskhalava:2011va}%
  \BibitemOpen
  \bibfield  {author} {\bibinfo {author} {\bibfnamefont {D.}~\bibnamefont
  {Pirtskhalava}}\ and\ \bibinfo {author} {\bibfnamefont {P.}~\bibnamefont
  {Uttayarat}},\ }\href {\doibase 10.1016/j.physletb.2012.04.039} {\bibfield
  {journal} {\bibinfo  {journal} {Phys. Lett.}\ }\textbf {\bibinfo {volume}
  {B712}},\ \bibinfo {pages} {81} (\bibinfo {year} {2012})},\ \Eprint
  {http://arxiv.org/abs/1112.5451} {arXiv:1112.5451 [hep-ph]} \BibitemShut
  {NoStop}%
\bibitem [{\citenamefont {Chau}\ and\ \citenamefont
  {Cheng}(1994)}]{Chau:1993ec}%
  \BibitemOpen
  \bibfield  {author} {\bibinfo {author} {\bibfnamefont {L.-L.}\ \bibnamefont
  {Chau}}\ and\ \bibinfo {author} {\bibfnamefont {H.-Y.}\ \bibnamefont
  {Cheng}},\ }\href {\doibase 10.1016/0370-2693(94)90176-7} {\bibfield
  {journal} {\bibinfo  {journal} {Phys. Lett.}\ }\textbf {\bibinfo {volume}
  {B333}},\ \bibinfo {pages} {514} (\bibinfo {year} {1994})},\ \Eprint
  {http://arxiv.org/abs/hep-ph/9404207} {arXiv:hep-ph/9404207 [hep-ph]}
  \BibitemShut {NoStop}%
\bibitem [{\citenamefont {Browder}\ and\ \citenamefont
  {Pakvasa}(1996)}]{Browder:1995ay}%
  \BibitemOpen
  \bibfield  {author} {\bibinfo {author} {\bibfnamefont {T.~E.}\ \bibnamefont
  {Browder}}\ and\ \bibinfo {author} {\bibfnamefont {S.}~\bibnamefont
  {Pakvasa}},\ }\href {\doibase 10.1016/0370-2693(96)00776-9} {\bibfield
  {journal} {\bibinfo  {journal} {Phys. Lett.}\ }\textbf {\bibinfo {volume}
  {B383}},\ \bibinfo {pages} {475} (\bibinfo {year} {1996})},\ \Eprint
  {http://arxiv.org/abs/hep-ph/9508362} {arXiv:hep-ph/9508362 [hep-ph]}
  \BibitemShut {NoStop}%
\bibitem [{\citenamefont {Wolfenstein}(1995)}]{Wolfenstein:1995kv}%
  \BibitemOpen
  \bibfield  {author} {\bibinfo {author} {\bibfnamefont {L.}~\bibnamefont
  {Wolfenstein}},\ }\href {\doibase 10.1103/PhysRevLett.75.2460} {\bibfield
  {journal} {\bibinfo  {journal} {Phys. Rev. Lett.}\ }\textbf {\bibinfo
  {volume} {75}},\ \bibinfo {pages} {2460} (\bibinfo {year} {1995})},\ \Eprint
  {http://arxiv.org/abs/hep-ph/9505285} {arXiv:hep-ph/9505285 [hep-ph]}
  \BibitemShut {NoStop}%
\bibitem [{\citenamefont {Blaylock}\ \emph {et~al.}(1995)\citenamefont
  {Blaylock}, \citenamefont {Seiden},\ and\ \citenamefont
  {Nir}}]{Blaylock:1995ay}%
  \BibitemOpen
  \bibfield  {author} {\bibinfo {author} {\bibfnamefont {G.}~\bibnamefont
  {Blaylock}}, \bibinfo {author} {\bibfnamefont {A.}~\bibnamefont {Seiden}}, \
  and\ \bibinfo {author} {\bibfnamefont {Y.}~\bibnamefont {Nir}},\ }\href
  {\doibase 10.1016/0370-2693(95)00787-L} {\bibfield  {journal} {\bibinfo
  {journal} {Phys. Lett.}\ }\textbf {\bibinfo {volume} {B355}},\ \bibinfo
  {pages} {555} (\bibinfo {year} {1995})},\ \Eprint
  {http://arxiv.org/abs/hep-ph/9504306} {arXiv:hep-ph/9504306 [hep-ph]}
  \BibitemShut {NoStop}%
\bibitem [{\citenamefont {Falk}\ \emph {et~al.}(1999)\citenamefont {Falk},
  \citenamefont {Nir},\ and\ \citenamefont {Petrov}}]{Falk:1999ts}%
  \BibitemOpen
  \bibfield  {author} {\bibinfo {author} {\bibfnamefont {A.~F.}\ \bibnamefont
  {Falk}}, \bibinfo {author} {\bibfnamefont {Y.}~\bibnamefont {Nir}}, \ and\
  \bibinfo {author} {\bibfnamefont {A.~A.}\ \bibnamefont {Petrov}},\ }\href
  {\doibase 10.1088/1126-6708/1999/12/019} {\bibfield  {journal} {\bibinfo
  {journal} {JHEP}\ }\textbf {\bibinfo {volume} {12}},\ \bibinfo {pages} {019}
  (\bibinfo {year} {1999})},\ \Eprint {http://arxiv.org/abs/hep-ph/9911369}
  {arXiv:hep-ph/9911369 [hep-ph]} \BibitemShut {NoStop}%
\bibitem [{\citenamefont {Gronau}\ and\ \citenamefont
  {Rosner}(2001)}]{Gronau:2000ru}%
  \BibitemOpen
  \bibfield  {author} {\bibinfo {author} {\bibfnamefont {M.}~\bibnamefont
  {Gronau}}\ and\ \bibinfo {author} {\bibfnamefont {J.~L.}\ \bibnamefont
  {Rosner}},\ }\href {\doibase 10.1016/S0370-2693(01)00080-6} {\bibfield
  {journal} {\bibinfo  {journal} {Phys. Lett.}\ }\textbf {\bibinfo {volume}
  {B500}},\ \bibinfo {pages} {247} (\bibinfo {year} {2001})},\ \Eprint
  {http://arxiv.org/abs/hep-ph/0010237} {arXiv:hep-ph/0010237 [hep-ph]}
  \BibitemShut {NoStop}%
\bibitem [{\citenamefont {Bergmann}\ \emph {et~al.}(2000)\citenamefont
  {Bergmann}, \citenamefont {Grossman}, \citenamefont {Ligeti}, \citenamefont
  {Nir},\ and\ \citenamefont {Petrov}}]{Bergmann:2000id}%
  \BibitemOpen
  \bibfield  {author} {\bibinfo {author} {\bibfnamefont {S.}~\bibnamefont
  {Bergmann}}, \bibinfo {author} {\bibfnamefont {Y.}~\bibnamefont {Grossman}},
  \bibinfo {author} {\bibfnamefont {Z.}~\bibnamefont {Ligeti}}, \bibinfo
  {author} {\bibfnamefont {Y.}~\bibnamefont {Nir}}, \ and\ \bibinfo {author}
  {\bibfnamefont {A.~A.}\ \bibnamefont {Petrov}},\ }\href {\doibase
  10.1016/S0370-2693(00)00772-3} {\bibfield  {journal} {\bibinfo  {journal}
  {Phys. Lett.}\ }\textbf {\bibinfo {volume} {B486}},\ \bibinfo {pages} {418}
  (\bibinfo {year} {2000})},\ \Eprint {http://arxiv.org/abs/hep-ph/0005181}
  {arXiv:hep-ph/0005181 [hep-ph]} \BibitemShut {NoStop}%
\bibitem [{\citenamefont {Falk}\ \emph {et~al.}(2002)\citenamefont {Falk},
  \citenamefont {Grossman}, \citenamefont {Ligeti},\ and\ \citenamefont
  {Petrov}}]{Falk:2001hx}%
  \BibitemOpen
  \bibfield  {author} {\bibinfo {author} {\bibfnamefont {A.~F.}\ \bibnamefont
  {Falk}}, \bibinfo {author} {\bibfnamefont {Y.}~\bibnamefont {Grossman}},
  \bibinfo {author} {\bibfnamefont {Z.}~\bibnamefont {Ligeti}}, \ and\ \bibinfo
  {author} {\bibfnamefont {A.~A.}\ \bibnamefont {Petrov}},\ }\href {\doibase
  10.1103/PhysRevD.65.054034} {\bibfield  {journal} {\bibinfo  {journal} {Phys.
  Rev.}\ }\textbf {\bibinfo {volume} {D65}},\ \bibinfo {pages} {054034}
  (\bibinfo {year} {2002})},\ \Eprint {http://arxiv.org/abs/hep-ph/0110317}
  {arXiv:hep-ph/0110317 [hep-ph]} \BibitemShut {NoStop}%
\bibitem [{\citenamefont {Kagan}\ and\ \citenamefont
  {Sokoloff}(2009)}]{Kagan:2009gb}%
  \BibitemOpen
  \bibfield  {author} {\bibinfo {author} {\bibfnamefont {A.~L.}\ \bibnamefont
  {Kagan}}\ and\ \bibinfo {author} {\bibfnamefont {M.~D.}\ \bibnamefont
  {Sokoloff}},\ }\href {\doibase 10.1103/PhysRevD.80.076008} {\bibfield
  {journal} {\bibinfo  {journal} {Phys. Rev.}\ }\textbf {\bibinfo {volume}
  {D80}},\ \bibinfo {pages} {076008} (\bibinfo {year} {2009})},\ \Eprint
  {http://arxiv.org/abs/0907.3917} {arXiv:0907.3917 [hep-ph]} \BibitemShut
  {NoStop}%
\bibitem [{\citenamefont {Aaij}\ \emph
  {et~al.}(2017{\natexlab{b}})\citenamefont {Aaij} \emph
  {et~al.}}]{Aaij:2016roz}%
  \BibitemOpen
  \bibfield  {author} {\bibinfo {author} {\bibfnamefont {R.}~\bibnamefont
  {Aaij}} \emph {et~al.} (\bibinfo {collaboration} {LHCb}),\ }\href {\doibase
  10.1103/PhysRevD.96.099907, 10.1103/PhysRevD.95.052004} {\bibfield  {journal}
  {\bibinfo  {journal} {Phys. Rev.}\ }\textbf {\bibinfo {volume} {D95}},\
  \bibinfo {pages} {052004} (\bibinfo {year} {2017}{\natexlab{b}})},\ \bibinfo
  {note} {[Erratum: Phys. Rev.D96,no.9,099907(2017)]},\ \Eprint
  {http://arxiv.org/abs/1611.06143} {arXiv:1611.06143 [hep-ex]} \BibitemShut
  {NoStop}%
\bibitem [{\citenamefont {Bigi}\ and\ \citenamefont
  {Sanda}(1986)}]{Bigi:1986dp}%
  \BibitemOpen
  \bibfield  {author} {\bibinfo {author} {\bibfnamefont {I.~I.~Y.}\
  \bibnamefont {Bigi}}\ and\ \bibinfo {author} {\bibfnamefont {A.~I.}\
  \bibnamefont {Sanda}},\ }\href {\doibase 10.1016/0370-2693(86)91557-1}
  {\bibfield  {journal} {\bibinfo  {journal} {Phys. Lett.}\ }\textbf {\bibinfo
  {volume} {B171}},\ \bibinfo {pages} {320} (\bibinfo {year}
  {1986})}\BibitemShut {NoStop}%
\bibitem [{\citenamefont {Xing}(1997)}]{Xing:1996pn}%
  \BibitemOpen
  \bibfield  {author} {\bibinfo {author} {\bibfnamefont {Z.-z.}\ \bibnamefont
  {Xing}},\ }\href {\doibase 10.1103/PhysRevD.55.196} {\bibfield  {journal}
  {\bibinfo  {journal} {Phys. Rev.}\ }\textbf {\bibinfo {volume} {D55}},\
  \bibinfo {pages} {196} (\bibinfo {year} {1997})},\ \Eprint
  {http://arxiv.org/abs/hep-ph/9606422} {arXiv:hep-ph/9606422 [hep-ph]}
  \BibitemShut {NoStop}%
\bibitem [{\citenamefont {Gronau}\ \emph {et~al.}(2001)\citenamefont {Gronau},
  \citenamefont {Grossman},\ and\ \citenamefont {Rosner}}]{Gronau:2001nr}%
  \BibitemOpen
  \bibfield  {author} {\bibinfo {author} {\bibfnamefont {M.}~\bibnamefont
  {Gronau}}, \bibinfo {author} {\bibfnamefont {Y.}~\bibnamefont {Grossman}}, \
  and\ \bibinfo {author} {\bibfnamefont {J.~L.}\ \bibnamefont {Rosner}},\
  }\href {\doibase 10.1016/S0370-2693(01)00426-9} {\bibfield  {journal}
  {\bibinfo  {journal} {Phys. Lett.}\ }\textbf {\bibinfo {volume} {B508}},\
  \bibinfo {pages} {37} (\bibinfo {year} {2001})},\ \Eprint
  {http://arxiv.org/abs/hep-ph/0103110} {arXiv:hep-ph/0103110 [hep-ph]}
  \BibitemShut {NoStop}%
\bibitem [{\citenamefont {Atwood}\ and\ \citenamefont
  {Petrov}(2005)}]{Atwood:2002ak}%
  \BibitemOpen
  \bibfield  {author} {\bibinfo {author} {\bibfnamefont {D.}~\bibnamefont
  {Atwood}}\ and\ \bibinfo {author} {\bibfnamefont {A.~A.}\ \bibnamefont
  {Petrov}},\ }\href {\doibase 10.1103/PhysRevD.71.054032} {\bibfield
  {journal} {\bibinfo  {journal} {Phys. Rev.}\ }\textbf {\bibinfo {volume}
  {D71}},\ \bibinfo {pages} {054032} (\bibinfo {year} {2005})},\ \Eprint
  {http://arxiv.org/abs/hep-ph/0207165} {arXiv:hep-ph/0207165 [hep-ph]}
  \BibitemShut {NoStop}%
\bibitem [{\citenamefont {Asner}\ and\ \citenamefont
  {Sun}(2006)}]{Asner:2005wf}%
  \BibitemOpen
  \bibfield  {author} {\bibinfo {author} {\bibfnamefont {D.~M.}\ \bibnamefont
  {Asner}}\ and\ \bibinfo {author} {\bibfnamefont {W.~M.}\ \bibnamefont
  {Sun}},\ }\href {\doibase 10.1103/PhysRevD.73.034024,
  10.1103/PhysRevD.77.019901} {\bibfield  {journal} {\bibinfo  {journal} {Phys.
  Rev.}\ }\textbf {\bibinfo {volume} {D73}},\ \bibinfo {pages} {034024}
  (\bibinfo {year} {2006})},\ \bibinfo {note} {[Erratum: Phys.
  Rev.D77,019901(2008)]},\ \Eprint {http://arxiv.org/abs/hep-ph/0507238}
  {arXiv:hep-ph/0507238 [hep-ph]} \BibitemShut {NoStop}%
\bibitem [{\citenamefont {Asner}\ \emph {et~al.}(2012)\citenamefont {Asner}
  \emph {et~al.}}]{Asner:2012xb}%
  \BibitemOpen
  \bibfield  {author} {\bibinfo {author} {\bibfnamefont {D.~M.}\ \bibnamefont
  {Asner}} \emph {et~al.} (\bibinfo {collaboration} {CLEO}),\ }\href {\doibase
  10.1103/PhysRevD.86.112001} {\bibfield  {journal} {\bibinfo  {journal} {Phys.
  Rev.}\ }\textbf {\bibinfo {volume} {D86}},\ \bibinfo {pages} {112001}
  (\bibinfo {year} {2012})},\ \Eprint {http://arxiv.org/abs/1210.0939}
  {arXiv:1210.0939 [hep-ex]} \BibitemShut {NoStop}%
\bibitem [{\citenamefont {Goldhaber}\ and\ \citenamefont
  {Rosner}(1977)}]{Goldhaber:1976fp}%
  \BibitemOpen
  \bibfield  {author} {\bibinfo {author} {\bibfnamefont {M.}~\bibnamefont
  {Goldhaber}}\ and\ \bibinfo {author} {\bibfnamefont {J.~L.}\ \bibnamefont
  {Rosner}},\ }\href {\doibase 10.1103/PhysRevD.15.1254} {\bibfield  {journal}
  {\bibinfo  {journal} {Phys. Rev.}\ }\textbf {\bibinfo {volume} {D15}},\
  \bibinfo {pages} {1254} (\bibinfo {year} {1977})}\BibitemShut {NoStop}%
\bibitem [{\citenamefont {Xing}(1996{\natexlab{a}})}]{Xing:1994mn}%
  \BibitemOpen
  \bibfield  {author} {\bibinfo {author} {\bibfnamefont {Z.-z.}\ \bibnamefont
  {Xing}},\ }\href {\doibase 10.1103/PhysRevD.53.204} {\bibfield  {journal}
  {\bibinfo  {journal} {Phys. Rev.}\ }\textbf {\bibinfo {volume} {D53}},\
  \bibinfo {pages} {204} (\bibinfo {year} {1996}{\natexlab{a}})},\ \Eprint
  {http://arxiv.org/abs/hep-ex/9502001} {arXiv:hep-ex/9502001 [hep-ex]}
  \BibitemShut {NoStop}%
\bibitem [{\citenamefont {Xing}(1996{\natexlab{b}})}]{Xing:1995vj}%
  \BibitemOpen
  \bibfield  {author} {\bibinfo {author} {\bibfnamefont {Z.-z.}\ \bibnamefont
  {Xing}},\ }\href {\doibase 10.1016/0370-2693(96)00077-9} {\bibfield
  {journal} {\bibinfo  {journal} {Phys. Lett.}\ }\textbf {\bibinfo {volume}
  {B372}},\ \bibinfo {pages} {317} (\bibinfo {year} {1996}{\natexlab{b}})},\
  \Eprint {http://arxiv.org/abs/hep-ph/9512216} {arXiv:hep-ph/9512216 [hep-ph]}
  \BibitemShut {NoStop}%
\bibitem [{\citenamefont {Xing}(1996{\natexlab{c}})}]{Xing:1995vn}%
  \BibitemOpen
  \bibfield  {author} {\bibinfo {author} {\bibfnamefont {Z.-z.}\ \bibnamefont
  {Xing}},\ }\href {\doibase 10.1016/0370-2693(96)00449-2} {\bibfield
  {journal} {\bibinfo  {journal} {Phys. Lett.}\ }\textbf {\bibinfo {volume}
  {B379}},\ \bibinfo {pages} {257} (\bibinfo {year} {1996}{\natexlab{c}})},\
  \Eprint {http://arxiv.org/abs/hep-ph/9512217} {arXiv:hep-ph/9512217 [hep-ph]}
  \BibitemShut {NoStop}%
\bibitem [{\citenamefont {Xing}(1999)}]{Xing:1999yw}%
  \BibitemOpen
  \bibfield  {author} {\bibinfo {author} {\bibfnamefont {Z.-z.}\ \bibnamefont
  {Xing}},\ }\href {\doibase 10.1016/S0370-2693(99)00978-8} {\bibfield
  {journal} {\bibinfo  {journal} {Phys. Lett.}\ }\textbf {\bibinfo {volume}
  {B463}},\ \bibinfo {pages} {323} (\bibinfo {year} {1999})},\ \Eprint
  {http://arxiv.org/abs/hep-ph/9907454} {arXiv:hep-ph/9907454 [hep-ph]}
  \BibitemShut {NoStop}%
\bibitem [{\citenamefont {Asner}\ \emph {et~al.}(2008)\citenamefont {Asner}
  \emph {et~al.}}]{Asner:2008ft}%
  \BibitemOpen
  \bibfield  {author} {\bibinfo {author} {\bibfnamefont {D.~M.}\ \bibnamefont
  {Asner}} \emph {et~al.} (\bibinfo {collaboration} {CLEO}),\ }\href {\doibase
  10.1103/PhysRevD.78.012001} {\bibfield  {journal} {\bibinfo  {journal} {Phys.
  Rev.}\ }\textbf {\bibinfo {volume} {D78}},\ \bibinfo {pages} {012001}
  (\bibinfo {year} {2008})},\ \Eprint {http://arxiv.org/abs/0802.2268}
  {arXiv:0802.2268 [hep-ex]} \BibitemShut {NoStop}%
\bibitem [{\citenamefont {Xing}(2019)}]{Xing:2019uzz}%
  \BibitemOpen
  \bibfield  {author} {\bibinfo {author} {\bibfnamefont {Z.-z.}\ \bibnamefont
  {Xing}},\ }\href@noop {} {\  (\bibinfo {year} {2019})},\ \Eprint
  {http://arxiv.org/abs/1903.09566} {arXiv:1903.09566 [hep-ph]} \BibitemShut
  {NoStop}%
\bibitem [{\citenamefont {Bigi}\ \emph {et~al.}(2011)\citenamefont {Bigi},
  \citenamefont {Paul},\ and\ \citenamefont {Recksiegel}}]{Bigi:2011re}%
  \BibitemOpen
  \bibfield  {author} {\bibinfo {author} {\bibfnamefont {I.~I.}\ \bibnamefont
  {Bigi}}, \bibinfo {author} {\bibfnamefont {A.}~\bibnamefont {Paul}}, \ and\
  \bibinfo {author} {\bibfnamefont {S.}~\bibnamefont {Recksiegel}},\ }\href
  {\doibase 10.1007/JHEP06(2011)089} {\bibfield  {journal} {\bibinfo  {journal}
  {JHEP}\ }\textbf {\bibinfo {volume} {06}},\ \bibinfo {pages} {089} (\bibinfo
  {year} {2011})},\ \Eprint {http://arxiv.org/abs/1103.5785} {arXiv:1103.5785
  [hep-ph]} \BibitemShut {NoStop}%
\bibitem [{\citenamefont {Buras}(1989)}]{Buras:1988ky}%
  \BibitemOpen
  \bibfield  {author} {\bibinfo {author} {\bibfnamefont {A.~J.}\ \bibnamefont
  {Buras}},\ }\bibfield  {booktitle} {\emph {\bibinfo {booktitle} {{XIX
  International Seminar on Theoretical Physics: Nonperturbative Aspects of the
  Standard Model (GIFT Seminar) Jaca (Huesca), Spain, June 6-11, 1988}}},\
  }\href {\doibase 10.1016/0920-5632(89)90066-2} {\bibfield  {journal}
  {\bibinfo  {journal} {Nucl. Phys. Proc. Suppl.}\ }\textbf {\bibinfo {volume}
  {10A}},\ \bibinfo {pages} {199} (\bibinfo {year} {1989})}\BibitemShut
  {NoStop}%
\bibitem [{\citenamefont {Hansen}\ and\ \citenamefont
  {Sharpe}(2012)}]{Hansen:2012tf}%
  \BibitemOpen
  \bibfield  {author} {\bibinfo {author} {\bibfnamefont {M.~T.}\ \bibnamefont
  {Hansen}}\ and\ \bibinfo {author} {\bibfnamefont {S.~R.}\ \bibnamefont
  {Sharpe}},\ }\href {\doibase 10.1103/PhysRevD.86.016007} {\bibfield
  {journal} {\bibinfo  {journal} {Phys. Rev.}\ }\textbf {\bibinfo {volume}
  {D86}},\ \bibinfo {pages} {016007} (\bibinfo {year} {2012})},\ \Eprint
  {http://arxiv.org/abs/1204.0826} {arXiv:1204.0826 [hep-lat]} \BibitemShut
  {NoStop}%
\end{thebibliography}%
\bibliographystyle{apsrev4-1}

\end{document}